\documentclass{article}
\usepackage{graphicx}
\usepackage{sidecap}
\usepackage{subfigure}
\usepackage{amssymb}
\usepackage{amsthm}
\usepackage{amsmath}
\usepackage{bm}             %% boldmath-command: \bm{}
\usepackage[mathscr]{eucal}
\usepackage{cancel}
\usepackage{psfrag}
\usepackage{wasysym}

\def\pmb#1{\setbox0=\hbox{$#1$}%
  \kern-.025em\copy0\kern-\wd0
  \kern.05em\copy0\kern-\wd0
  \kern-.025em\raise.0433em\box0}
\def\pmbs#1{\setbox0=\hbox{$\scriptstyle #1$}%
  \kern-.0175em\copy0\kern-\wd0
  \kern.035em\copy0\kern-\wd0
  \kern-.0175em\raise.0303em\box0}
\def\be{\begin{equation}}
\def\ee{\end{equation}}
\def\bea{\begin{eqnarray}}
\def\eea{\end{eqnarray}}
\def\lb{\label}

\def\bi{\bibitem}
\def\vec#1{\mbox{\boldmath$#1$}}

\def\gam{\gamma}
\def\d{\delta}
\def\eps{\epsilon}

\def\sig{\sigma}

\def\Sig{\Sigma}

\def\Om{\Omega}

\def\bom{\mbox{\boldmath $\omega$}}

\def\bna{\mbox{\boldmath $\nabla$}}

\def\vece{\vec{e}}

\def\ptl{\partial}

\def\la{\langle}
\def\ra{\rangle}

\def\hsp5{\hspace{5mm}}
\def\case#1/#2{\textstyle\frac{#1}{#2}}
\newcommand{\sfrac}[2]{{\textstyle{#1\over#2}}}
\theoremstyle{plain}
\newtheorem{theorem}{Theorem}[section]

\newtheorem{lemma}[theorem]{Lemma}

\newtheorem{conjecture}[theorem]{Conjecture}

\theoremstyle{remark}

\newtheorem*{remark}{Remark}
\title{\sc Bianchi type I models with two tilted fluids}
\author{\sc
 Patrik Sandin $^{1}$\thanks{Electronic address: {\tt patrik.sandin@kau.se}}\
,\ \ and Claes Uggla$^{1}$\thanks{Electronic address:
{\tt claes.uggla@kau.se}}\\
$^{1}${\small\em Department of Physics, University of Karlstad,}\\
{\small\em S-651 88 Karlstad, Sweden}}

\begin{document}
%%%%%%%%%%%%%%%%%%%%%%%%%%%%%%%%%%%%%%%%%%%%%%%%%%%%%%%%%%%%%%%%%%%
\maketitle
%\sloppy
%\doublespace

%%%%%%%%%%%%%%%%%%%%%%%%%%%%%%%%%%%%%%%%%%%%%%%%%%%%%%%%%%%%%%%%%%%
\begin{abstract}
%%%%%%%%%%%%%%%%%%%%%%%%%%%%%%%%%%%%%%%%%%%%%%%%%%%%%%%%%%%%%%%%%%%

In this paper we investigate expanding Bianchi type I models
with two tilted fluids with linear equations of state.
Individually the fluids have non-zero energy fluxes w.r.t. the
symmetry surfaces, but these cancel each other because of the
Codazzi constraint. Asymptotically toward the past the
solutions approach Kasner states if the speeds of sound are
less than that of light. If one of the fluids has a speed of
sound that is less or equal to $1/3$ of the speed of light
(radiation) then the models isotropize toward the future, but
if both fluids are stiffer than radiation then the final state
is anisotropic with non-zero Hubble-normalized shear. The
significance of these results is discussed in a broader
context.

%%%%%%%%%%%%%%%%%%%%%%%%%%%%%%%%%%%%%%%%%%%%%%%%%%%%%%%%%%%%%%%%%%%
\end{abstract}
%%%%%%%%%%%%%%%%%%%%%%%%%%%%%%%%%%%%%%%%%%%%%%%%%%%%%%%%%%%%%%%%%%%

\centerline{\bigskip\noindent PACS numbers: 04.20.-q, 04.20.Dw,
04.20.Ha, 98.80.-k, 98.80.Bp, 98.80.Jk} \vfill
\newpage

%%%%%%%%%%%%%%%%%%%%%%%%%%%%%%%%%%%%%%%%%%%%%%%%%%%%%%%%%%%%%%%%%%%
\section{Introduction}\label{Sec:intro}
%%%%%%%%%%%%%%%%%%%%%%%%%%%%%%%%%%%%%%%%%%%%%%%%%%%%%%%%%%%%%%%%%%%

The construction of a relativistic model of gravity contains
the following ingredients: (i) a 4-dimensional manifold ${\cal
M}$ endowed with a Lorentzian metric, (ii) a matter source
description, (iii) dynamical laws---Einstein's field equations
and, if needed, matter equations. A general relativistic {\em
cosmological\/} model requires that one in addition attempts to
describe the universe at a particular scale.

At the largest spatial scales present observational data
suggest that the `standard' $\Lambda{\rm CDM}$ model of
cosmology provides the most simple consistent description of
the universe today. This model is spatially homogeneous (SH)
and isotropic with flat spatial geometry; the matter content
consists of a dark energy component, modeled by a positive
cosmological constant, supplemented with dark matter and atoms,
described by pressureless fluids with $n^a$, the unit normal to
the symmetry surfaces, as the common 4-velocity. Density
fluctuations, described by linear scalar perturbations, are
seeded by an almost Gaussian, adiabatic, nearly scale invariant
process, see e.g.~\cite{hinetal08},\cite{tegetal06}, and
references therein.

However, this description does not hold on all scales, neither
spatial nor temporal. On smaller spatial scales matter has to
be described by many components, with energy fluxes in
different directions, e.g., our galaxy is moving w.r.t. to the
CMB. In the very early universe, and perhaps also in the
distant future, the $\Lambda{\rm CDM}$ model does not give a
correct matter description, indeed, although radiation can be
treated as a gravitational test field\footnote{A gravitational
test field is a field that does not affect the metric, i.e., it
is a field for which we can neglect: (i) its source
contribution to Einstein's equations, (ii) its influence on
gravitational sources.} today it has been an important
gravitational source in the past, and inflationary proponents
suggest that matter could be extremely different in the very
early universe. Clearly the matter description in the
`standard' scenario is not exact, and even with a few matter
components the associated energy fluxes cannot all be {\em
exactly\/} aligned, not even on the largest spatial scales.
What then happens with the matter and its associated energy
fluxes in the far future and what was the situation in the
distant past?

We believe that we understand how radiation and matter
interacts, at least after the very early universe when we think
our empirical experience holds. Presumably this interaction
explains why the 4-velocities associated with radiation and
matter today are fairly well aligned with each other on large
spatial scales, or maybe this alignment was produced in the
very early universe by some unknown process, perhaps inflation.
But is it obvious that this alignment should have persisted to
the extent present observations indicate after recombination,
and is it going to persist in the far future? Do linear vector
perturbations of SH and isotropic models suffice to determine
this? In the early universe interactions presumably played an
important role in aligning energy fluxes of different matter
components, but it is unclear what those interactions were; is
it possible to shed any light on this issue without knowing the
details of these interactions?

Here we are going to consider two non-interacting perfect
fluids that in general move w.r.t. each other and a
non-negative cosmological constant, which includes the
$\Lambda{\rm CDM}$ matter content as a special case. Although
this may not be a good matter description at all times, it is
still a useful step since it allows a comparative study of the
effects of various types of interactions in possible future
projects, an issue we return to in the concluding remarks.

Two matter source components with energy fluxes in different
directions produce an anisotropic source which excludes the
isotropic standard model and forces one to consider anisotropic
geometries. There are several reasons that suggest that the
natural anisotropic models to start with are the SH Bianchi
type I models. One reason is their geometric simplicity since
this sheds light on more general models---if type I turns out
to be complicated, then more general models will be even worse.
But more importantly is that they are the foundation in a
hierarchy of ever more geometrically complex models.

The SH Bianchi models (models that admit simply transitive
3-dimensional symmetry groups) form a crucial level in the
geometric complexity hierarchy, and within this level the
Bianchi type I models is the common ingredient since they can
be obtained from all other Bianchi models by Lie algebra
contractions. The consequences of this property are revealed
when one casts Einstein's equations into a dynamical system
where the type I models appear as part of a state space
boundary that describes asymptotic features of all other
Bianchi models, see e.g.~\cite{waiell97} and~\cite{col03}, and
references therein.

Moreover, the Bianchi models themselves serve as building
blocks for understanding asymptotic dynamics of more general
inhomogeneous models, the primary reason being the following:
In the very early universe near a generic or isotropic
singularity, or in the very late universe in an inflationary
epoch, horizons form and asymptotically shrink, asymptotically
prohibiting communication---a phenomenon naturally referred to
as asymptotic silence---generically pushing inhomogeneities
outside the horizons leading to that the asymptotic evolution
can locally be described by SH models---asymptotic locality,
see~\cite{uggetal03},\cite{andetal05},\cite{heietal07}. In the
dynamical systems approach these features are formally captured
by recasting the field equations into an infinite dimensional
dynamical system that at each spatial point has a
boundary---the silent boundary, which asymptotically attracts
generic asymptotically local dynamics. The dynamics on the
silent boundary, which determine the generic asymptotically
local dynamics, is described by a finite dimensional dynamical
system that is identical to that of the Bianchi models and
hence type I is a common key ingredient in a very general
context~\cite{uggetal03},\cite{andetal05},\cite{heietal07},\cite{rohugg05}.
Furthermore, there are hints, analytical and numerical, that
Bianchi models are important for describing future asymptotic
states, even in the absence of inflation and asymptotic silence
and locality, that sheds light on spatial structure formation.

This is not the first study using dynamical systems techniques
for studying multi-fluid models. Bianchi models with two fluids
with both 4-velocities being orthogonal to the symmetry
surfaces were studied in~\cite{colwai92}. However, it is not
difficult to predict what is going to happen when the
non-interacting fluids are aligned with each other, even in the
general inhomogeneous case. Let us introduce a length scale
$\ell$ defined by $\dot{\ell}/\ell=H$, where $H$ is the Hubble
scalar and where the dot refers to the time derivative w.r.t.
the proper clock time along the common fluid congruence. For
simplicity we consider fluids with linear equations of state
such that $p=w\rho$, where $p$, $\rho$ is the pressure and
energy density, respectively, and where the constant $w$
describes the speed of sound $c_s$ according to $w=c_s^2$ when
$w\geq 0$. Interesting examples of equations of state are:
$w=-1$, which corresponds to a positive cosmological constant;
dust, $w=0$; radiation, $w=\frac{1}{3}$; and a stiff fluid,
$w=1$, for which the speed of sound is equal to that of light.
Local energy-momentum conservation then yields that
$\rho\propto \ell^{-3(1+w)}$, see e.g.~\cite{waiell97},
\cite{heietal05}. Thus if $\ell\rightarrow 0$ the energy
densities of fluids with smaller $w$ become asymptotically
negligible compared to those with larger $w$, while if
$\ell\rightarrow \infty$ the opposite holds.

However, the situation is much more complicated when the
non-interacting fluids are not aligned. In~\cite{golnil00}
Bianchi type V models with two fluids and a positive
cosmological constant were studied; one of the fluids had a
flow orthogonal to the symmetry surfaces while the other had a
`tilted' flow, i.e., its 4-velocity was not aligned with the
normal to the SH surfaces. In~\cite{colher04}, where
brane-world cosmology was invoked to motivate the study of
multiple fluids, Bianchi type VI$_0$ with two non-interacting
tilted fluids was investigated. Perhaps because the focus in
these papers was on the quite interesting late time behavior,
there was no mentioning about the possibility of having two (or
more) tilted fluids in Bianchi type I.

The present paper is organized as follows: In the next section
we derive a reduced dynamical system that describes Bianchi
type I with two non-interacting fluids with linear equation of
state and a cosmological constant. In the subsequent section we
describe the associated state space and the influence of a
positive cosmological constant; in addition we list the
invariant subsets and fix points that are essential for
understanding the present models. In section~\ref{Sec:mon} we
give several monotonic functions that are useful for
determining the asymptotic dynamics; we also briefly discuss
some reasons why they exist since this allows one to produce
monotonic functions for other models as well.
Section~\ref{Sec:attractor} takes the results in the previous
sections as the starting point for a dynamical systems analysis
which yield our main results about asymptotic dynamics toward
the past and toward the future, in the absence of a
cosmological constant. We conclude with a discussion in
section~\ref{Sec:concl} about the significance of our results
in a more general context. Appendix~\ref{Sec:locstab} contains
detailed information about the fix points and their stability
properties. Throughout we use units such that $c=1=8 \pi G$.

%%%%%%%%%%%%%%%%%%%%%%%%%%%%%%%%%%%%%%%%%%%%%%%%%%%%%%%%%%%%%%%%%%%%
\section{Derivation of the dynamical system}
\label{Sec:dynsysder}
%%%%%%%%%%%%%%%%%%%%%%%%%%%%%%%%%%%%%%%%%%%%%%%%%%%%%%%%%%%%%%%%%%%%

%---------------------------------------------------------------------------------
%\subsection*{Bianchi models with an arbitrary source} \label{BIder}
%---------------------------------------------------------------------------------

In the orthonormal frame approach one uses a tetrad of four
orthogonal unit basis vector fields $\{\,{\bf e}_a\,\}$ and the
associated dual one-forms $\{\,{\bom}^a\,\}$ ($a=0,1,2,3$),
which, when expressed in a local coordinate basis, take the
form ${\bf e}_a = e_a{}^\mu\ptl/\ptl x^\mu\ =
e_a{}^\mu\ptl_\mu$,\, $\bom^a = e^a{}_\mu\,{\bf d}x^\mu$
($\mu=0,1,2,3$), where the tetrad components $e_a{}^\mu
(x^\nu)$ and their inverse components $e^a{}_\mu (x^\nu)$
satisfy the duality relations $e_a{}^\mu\,e^a{}_\nu =
\delta^\mu{}_\nu \,\Leftrightarrow \, e_a{}^\mu\,e^b{}_\mu =
\delta^b{}_a$, and where the orthogonality conditions are given
by $g_{ab} = {\bf e}_a\cdot {\bf e}_b =
g_{\mu\nu}\,e_a{}^\mu\,e_b{}^\nu = \eta_{ab}$;\, $g_{\mu\nu} =
\eta_{ab}\,e^a{}_\mu\,e^b{}_\nu$,\, $\eta_{ab} = {\rm
diag}\,(\,-1,1,1,1\,)$. The commutator functions
$c^a{}_{bc}(x^\mu)$, defined as $[{\bf e}_a,{\bf e}_b] =
c^c{}_{ab}\,{\bf e}_c$, are typically `elevated' to dependent
variables satisfying the Jacobi identities, ${\bf
e}_{[a}\,c^d{}_{bc]} - c^d{}_{e[a}\,c^e{}_{bc]} = 0$.

Let us now consider SH Bianchi models, i.e., models with a
foliation of SH hypersurfaces invariant under a simply
transitive group action $G_3$, and let us also introduce an
orthonormal basis of vector fields $\{\mathbf{e}_a\}$ that is
invariant under the group action such that the timelines are
orthogonal to the SH hypersurfaces with ${\bf e}_0={\bf n} =
\ptl/\ptl t$, where $t$ is the proper time along the geodesic
timelines (the geodesic property follows from the symmetries).
This yields the line-element: $ds^2 = - dt^2 +
\delta_{\alpha\beta}\,\bom^\alpha\otimes \bom^\beta$\,
($\alpha,\beta =1,2,3$), where $\bom^\alpha$ (with components
$e^\alpha{}_i$) are the one-forms dual to the triad
$\vece_\alpha$ (with components $e_\alpha{}^i$), tangential to
the symmetry surfaces, i.e., $e_\alpha{}^i e^\beta{}_i =
\delta^\alpha{}_\beta$ ($i=1,2,3$).

A 3+1 split of the commutator equations w.r.t. ${\bf e}_0={\bf
n}$ yields:
\begin{subequations}
\begin{align} \lb{dcomts0} [\,\vece_{0}, \vece_{\alpha}\,] &= -
[\,H\,\d_{\alpha}{}^{\beta} + \sig_{\alpha}{}^{\beta} +
\eps_{\alpha}{}^{\beta}{}_{\gam}\,\Omega^{\gam}\,]\,
\vece_{\beta}\:, \\
\lb{dcomtsa} [\,\vece_{\alpha}, \vece_{\beta}\,] &=
c^\gam{}_{\alpha\beta}\,\vece_{\gam} =
2a_{[\alpha}\,\d_{\beta]}{}^{\gam} +
\eps_{\alpha\beta\delta}\,n^{\delta\gam}\:.
\end{align}
\end{subequations}
Here $H$ is the Hubble variable, which is related to the
expansion $\theta$ of ${\bf n}$ according to
$H=\frac{1}{3}\theta$; $\sigma_{\alpha\beta}$ is the shear
associated with ${\bf n}$; $\Omega^\alpha$ is the Fermi
rotation which describes how the spatial triad rotates with
respect to a gyroscopically fixed so-called Fermi
frame;\footnote{The sign in the definition of $\Omega^\alpha$
is the same as in~\cite{ellels99}, but opposite of that
in~\cite{waiell97}.} $n^{\alpha\beta}$ and $a_\alpha$ describe
the Lie algebra of the 3-dimensional simply transitive Lie
group and determine the spatial three-curvature, see
e.g.~\cite{waiell97}.

Due to the symmetries $e_\alpha{}^i$ can be written as
$e_\alpha{}^i=\tilde{e}_\alpha{}^\beta(t)\hat{e}_\beta{}^i$,
where $\hat{e}_\alpha{}^i$ are functions of the spatial
coordinates $x^i$ alone such that $[\,\hat{\vece}_{\alpha},
\hat{\vece}_{\beta}\,] =
\hat{c}^\gam{}_{\alpha\beta}\,\hat{\vece}_{\gam} =
2\hat{a}_{[\alpha}\,\d_{\beta]}{}^{\gam} +
\eps_{\alpha\beta\delta}\,\hat{n}^{\delta\gam}$, where
$\hat{\vece}_\alpha = \hat{e}_\alpha{}^i\partial/\partial x^i$,
and where $\hat{c}^\gam{}_{\alpha\beta}$, parameterized by
$\hat{a}_\alpha , \hat{n}^{\alpha\beta}$, are the structure
constants of the symmetry group. The symmetries lead to that
the equations for the variables $\tilde{e}_\alpha{}^\beta(t)$
($d{\tilde{e}}_\alpha{}^\beta(t)/dt =
-[\,H\,\d_{\alpha}{}^{\gam} + \sig_{\alpha}{}^{\gam} +
\eps_{\alpha}{}^{\gam}{}_{\delta}\,\Omega^{\delta}\,]\tilde{e}_\gam{}^\beta$,
as follows from~\eqref{dcomts0}) decouple from the remaining
field equations, and because of this they are not usually
considered when discussing Bianchi models from an orthonormal
frame perspective.

A 3+1 split of the total stress-energy tensor $T_{ab}$ w.r.t.
$n^a$ yields:
\begin{subequations}\label{Tirreduc}
\begin{align}
T_{ab} &= \rho\,n_a\,n_b + 2q_{(a}\,n_{b)} + p\,h_{ab} + \pi_{ab}\:,\\
\rho &= n^a\,n^b\,T_{ab}\:,\qquad q_a = -h_a{}^b\,n^c\,T_{bc}\:,\qquad
p = \sfrac{1}{3}\,h^{ab}\,T_{ab}\:,\qquad \pi_{ab} = h_{\la
a}{}^c\,h_{b\ra}{}^d\,T_{cd}\:,
\end{align}
\end{subequations}
where $h_{ab} = n_a n_b + g_{ab}$; $\rho, p$ is the total
energy density and total effective pressure, respectively,
measured in the rest space of $n^a$; $\la..\ra$ stands for the
trace-free part of a symmetric spatial tensor, i.e. $A_{\la
\alpha\beta \ra}=A_{\alpha\beta} -
\frac{1}{3}\delta_{\alpha\beta}\,A^\gamma{}_\gamma$. In general
$T_{ab}$ consists of several components $T^{(i)}_{ab}$, such
that $T_{ab} = \sum_i T^{(i)}_{ab}$. If the components are
non-interacting, then $\bna_a T_{(i)}^{ab}=0$. A cosmological
constant $\Lambda$ can be formally regarded as a component of
$T_{ab}$ such that $\rho_\Lambda=\Lambda,\, p_\Lambda =
-\Lambda$, while $q_{(\Lambda)}^\alpha=0,\,
\pi_{(\Lambda)}^{\alpha\beta}=0$.

In the Hubble normalized approach one factors out the Hubble
variable $H$ by means of a conformal transformation which
yields dimensionless quantities~\cite{rohugg05}. In the
present SH case this amounts to the following:
\be (\Sigma_{\alpha\beta}, R^\alpha, N^{\alpha\beta},A_\alpha)
= \frac{1}{H}(\sigma_{\alpha\beta}, \Omega^\alpha,
n^{\alpha\beta},a_\alpha)\:,\qquad (\Omega, P, Q_\alpha,
\Pi_{\alpha\beta}) =
\frac{1}{3H^2}(\rho,p,q_\alpha,\pi_{\alpha\beta})\:,\ee
where we have chosen to normalize the stress-energy quantities
with $3H^2$ rather than $H^2$ in order to conform with the
usual definition of $\Omega$. In addition to this we choose
a new dimensionless time variable $\tau$ according to
\be \frac{d\tau}{dt} = H\:. \ee

Since $H$ is the only variable with dimension, its evolution
equation decouples from the remaining equations for dimensional
reasons:
\be\lb{Heq} H^\prime = -(1+q)H\:;\qquad\qquad  q = 2\Sig^{2} +
\sfrac{1}{2}(\Om+3P)\:,\qquad
\Sigma^2=\sfrac{1}{6}\Sigma_{\alpha\beta}\Sigma^{\alpha\beta}\:,
\ee
where a prime henceforth denotes $d/d\tau$ and where $q$ is the
deceleration parameter, obtained by means of one of Einstein's
equations---the Raychaudhuri equation; note that $\Omega$ and
$P$ in the expression for $q$ refers to the total
Hubble-normalized stress-energy content. A 3+1 split of the
remaining Einstein's field equations ($G_{ab}=T_{ab}$, where
$G_{ab}$ is the Einstein tensor and $T_{ab}$ the total
stress-energy tensor) and the Jacobi identities, yields the
following reduced system of coupled equations for the
Hubble-normalized variables:

\begin{subequations}\lb{devoleq}
\begin{align}
\Sigma_{\alpha\beta}^\prime &= -(2-q)\Sigma_{\alpha\beta} +
2\epsilon^{\gamma\delta}{}_{\la \alpha}\,\Sigma_{\beta\ra
\delta}\,R_\gamma - \,^3{\cal R}_{\la\alpha\beta\ra} + 3\Pi_{\alpha\beta}\:,\lb{HspatE}\\
A_{\alpha}^\prime &= [q\,\d_{\alpha}{}^{\beta} -
\Sig_{\alpha}{}^{\beta} - \eps_{\alpha}{}^{\beta}{}_{\gam}\,R^{\gam}] A_\beta\:,\lb{Hajac}\\
(N^{\alpha\beta})^\prime &= [q\,\d_{\gamma}{}^{(\alpha} +
2\Sig_{\gam}{}^{(\alpha} + 2
\eps_{\gam}{}^{(\alpha}{}_{\delta}\,R^{\delta}] N^{\beta )\gamma}\lb{Hnjac}\:,\\
0 &= 1 - \Sigma^2 + \sfrac{1}{6}\,^3{\cal R} - \Omega\:,\label{dGauss}\\
0 &= (3\delta_\alpha{}^\gamma\,A_\beta +
\epsilon_\alpha{}^{\delta\gamma}
\,N_{\delta\beta})\,\Sigma^\beta{}_\gamma - 3Q_\alpha\:,\label{dCodazzi}\\
0 &= A_\beta\, N^\beta{}_\alpha\:,\lb{HJacobi}
\end{align}
\end{subequations}
where $^3{\cal R}_{\la\alpha\beta\ra}$ and $^3{\cal R}$
describe the trace-free and scalar parts of the
Hubble-normalized three-curvature, respectively, according to:
\be\lb{threecurv} ^3{\cal R}_{\la\alpha\beta\ra} = B_{\la
\alpha\beta \ra} + 2\epsilon^{\gamma\delta}{}_{\la
\alpha}\,N_{\beta\ra\delta}\,A_\gamma\:,\quad ^3{\cal R} =
-\sfrac{1}{2}B^\alpha{}_\alpha - 6A^2\:;\qquad B_{\alpha\beta}
= 2 N_{\alpha\gamma}\,N^\gamma{}_\beta -
N^\gamma{}_\gamma\,N_{\alpha\beta}\:;
\ee
\eqref{HspatE} are the Hubble-normalized spatial and trace-free
Einstein equations; \eqref{Hajac} and~\eqref{Hnjac} are
evolution equations obtained from the Jacobi identities;
\eqref{dGauss} and \eqref{dCodazzi} are the Hubble-normalized
Gauss and Codazzi constraints, respectively, while the
constraint~\eqref{HJacobi} stems from the Jacobi identities.
The conservation law $\bna_a T^{ab}=0$ for the total
stress-energy tensor yields:
\begin{subequations}\lb{dmattereq}
\begin{align}
\lb{dlomdot} \Om^\prime &=  (2q-1)\,\Om - 3P +
2A_{\alpha}\,Q^{\alpha} - \Sig_{\alpha\beta}\Pi^{\alpha\beta}\:,\\
\lb{dlqmalpha} Q_{\alpha}^\prime &= -[2(1-q)\,\d_{\alpha}{}^{\beta}
+ \Sig_{\alpha}{}^{\beta} +
\eps_{\alpha}{}^{\beta}{}_{\gam}\,R^{\gam}]\,Q_{\beta} +
(3A_\beta\,\delta_\alpha{}^\delta -
\epsilon_{\alpha\beta}{}^\gamma\,N_\gamma{}^\delta)\,\Pi_\delta{}^\beta\:.
\end{align}
\end{subequations}

Let us now restrict ourselves to the {\em Bianchi type I\/} case with
expansion ($H>0$), characterized by
\be A_\alpha=0\:,\qquad N^{\alpha\beta}=0\:. \ee
In type I~\eqref{dlqmalpha} reduces to $Q_{\alpha}^\prime =
-[2(1-q)\,\d_{\alpha}{}^{\beta} + \Sig_{\alpha}{}^{\beta} +
\eps_{\alpha}{}^{\beta}{}_{\gam}\,R^{\gam}]\,Q_{\beta}$, and
hence the reduction of the Codazzi constraint~\eqref{dCodazzi}
to $Q_\alpha=0$ is consistent since it is preserved during
evolution. Thus there is no total energy flux in Bianchi type
I, and hence the type I SH frame is an energy frame, in the
nomenclature of Landau and Lifshitz~\cite{lanlif63}. Even so, a
matter source can consist of several components that
individually have non-zero energy fluxes, as long as they add
up to zero.

Let us now specialize to a source that consists of a
non-negative {\em cosmological constant\/} $\Lambda\geq 0$ {\em
and two non-interacting perfect fluids\/}, i.e., $T^{ab}_{(i)}
= (\tilde{\rho}_{(i)} + \tilde{p}_{(i)})
\tilde{u}^a_{(i)}\tilde{u}^b_{(i)} + \tilde{p}_{(i)} g^{ab}$;
$\bna_a T^{ab}_{(i)}=0$ ($i=1,2$), where $\tilde{\rho}_{(i)},
\tilde{p}_{(i)}$, is the energy density and pressure,
respectively, in the rest frame of the $i$:th fluid, while
$\tilde{u}^a_{(i)}$ is its 4-velocity. We assume that
$\tilde{\rho}_{(i)}\geq 0$, and for simplicity also a {\em
linear equations of state\/}, $\tilde{p}_{(i)} = w_{(i)}
\tilde{\rho}_{(i)}$, where $w_{(i)}=const$. The most
interesting equations of state are dust, $w=0$, and radiation,
$w=\frac{1}{3}$, but it is useful to not restrict oneself to
these values in order to study structural stability, however,
we do restrict ourselves to
\be 0 \leq w_{(2)} < w_{(1)} <1\:; \ee
since $w_{(1)}=1$, $w_{(1)}=w_{(2)}$ are associated with
bifurcations that needs special treatment, to be dealt with
elsewhere.\footnote{We could have extended the range of the
equation of state to include $-1/3<w_{(i)}<0$, but the
well-posedness of the Einstein equations for this range has
been questioned, see~\cite{friren00}.}

Making a 3+1 split with respect to ${\bf n}= \vece_0$ yields
\bea \tilde{u}^a_{(i)} = \Gamma_{(i)}(n^a + v^a_{(i)})\:;
\qquad n_a v^a_{(i)}=0\:,\qquad \Gamma_{(i)} = 1/\sqrt{1 -
v^2_{(i)}}\: \qquad (v^2_{(i)}=
\delta_{\alpha\beta}v_{(i)}^{\alpha}v_{(i)}^\beta)\:, \eea
which gives
\be\label{pfrel} Q_{(i)}^\alpha = (1 + w_{(i)})
(G^{(i)}_+)^{-1}\, v_{(i)}^\alpha\,\Omega_{(i)}\:,\quad\,\,
P_{(i)} = w_{(i)}\Omega_{(i)} + \sfrac{1}{3} (1 -
3w_{(i)})Q^{(i)}_\alpha v_{(i)}^\alpha\:,\quad\,\,
\Pi^{(i)}_{\alpha\beta} =
Q^{(i)}_{\la\alpha}v^{(i)}_{\beta\ra}\:, \ee
where $G^{(i)}_\pm  =  1 \pm w_{(i)} \, v_{(i)}^2$;
$\Omega_{(i)}=\rho_{(i)}/(3H^2)$. The cosmological constant
contributes $\Omega_\Lambda = \Lambda/(3H^2)=-P_\Lambda$ to the
total $\Omega$ and $P$, while
$Q^\alpha_\Lambda=0=\Pi^{\alpha\beta}_\Lambda$. Due to its
definition and equation~\eqref{Heq}, $\Omega_\Lambda$ satisfies
\be \Omega_\Lambda^\prime = 2(1+q)\Omega_\Lambda\:. \ee

The Codazzi constraint, $Q_\alpha= Q_\alpha^{(1)} +
Q_\alpha^{(2)}= 0$, taken in combination with~\eqref{pfrel}
forces the two fluids 3-velocities to be anti-parallel.
Kinematically the situation is similar to that of Bianchi type
I with a general magnetic field studied in~\cite{leb97}, and it
is therefore natural to exploit the same mathematical
structures in the present problem. We therefore choose the
spatial triad so that one of the frame vectors is aligned with
the fluid velocities, which we choose to be $\vece_3$, in
agreement with what is usually done in physics, i.e.
$v_{(i)}^{\alpha}= (0,0,v_{(i)})$. Demanding that these
conditions on $v_{(i)}^{\alpha}$ hold for all times lead to the
following conditions\footnote{In the case of a magnetic field,
aligned along $\vece_3$, one obtains $R_1 = \Sigma_{23},\,R_2 =
-\Sigma_{31}$, i.e., the signs are opposite of those of the two
tilted fluid case! This dynamical result in turn leads to sign
differences in the $\Sigma$-equations. Note that the kinematic
results in~\cite{leb97} still hold, hence e.g. fix points
correspond to transitively self-similar models.}
\be R_1 = -\Sigma_{23}\:,\qquad R_2 = \Sigma_{31}\:. \ee
This leaves $R_3$ undetermined, however, we still have the
freedom of arbitrary rotations in the 1-2-plane, which we use
to set
\be R_3=0\:.\ee
Following~\cite{leb97}, we introduce the variables
$\Sigma_+,\Sigma_A,\Sigma_B,\Sigma_C$ according to
\be \Sigma_+ = \sfrac{1}{2}(\Sigma_{11}+\Sigma_{22})\:,\qquad
\Sigma_{31} + \mathrm{i}\,\Sigma_{23} =  \sqrt{3}\Sigma_A\,
e^{\mathrm{i}\phi} \:,\qquad  \Sigma_-  +
\frac{\mathrm{i}}{\sqrt{3}} \Sigma_{12}= (\Sigma_B +
\mathrm{i}\,\Sigma_C) e^{2\mathrm{i}\phi}\:, \ee
where $\Sigma_-=(\Sigma_{11}-\Sigma_{22})/(2\sqrt{3})$, which
leads to
\be \Sigma^2= \Sigma_+^2 + \Sigma_A^2 + \Sigma_B^2 +
\Sigma_C^2\:. \ee

The above decomposition of $\Sigma_{\alpha\beta}$ has the
advantage that the equation for $\phi$, $d\phi/d\tau =
-\Sigma_C$, decouples from the other equations, leaving the
following reduced constrained dynamical system of coupled
equations for the Hubble-normalized shear variables
$\Sigma_+,\Sigma_A,\Sigma_B,\Sigma_C$, the fluid 3-velocities
$v_{(1)}, v_{(2)}$, the Hubble-normalized energy densities
$\Omega_{(1)},\Omega_{(2)}$, and the Hubble-normalized
cosmological constant $\Omega_\Lambda$:

\vspace*{2mm} \noindent {\em Evolution equations}:
\begin{subequations} \label{evolBI}
\begin{align}
\Sigma_+^\prime &= -(2-q)\Sigma_+ + 3\Sigma_A^2 -  Q_{(1)}v_{(1)} -  Q_{(2)}v_{(2)}\:, \label{Sigp}\\
\Sigma_A^\prime &= -(2 - q + 3\Sigma_+ + \sqrt{3}\Sigma_B) \Sigma_A \:, \label{SigA}\\
\Sigma_B^\prime &= -(2-q)\Sigma_B + \sqrt{3}\Sigma_A^2 -
2\sqrt{3}\Sigma_C^2\:, \\
\Sigma_C^\prime &= -(2 - q - 2\sqrt{3}\Sigma_B) \Sigma_C\:, \label{SigC}\\
v_{(i)}^\prime &= (G^{(i)}_-)^{-1} (1-v_{(i)}^2)(3w_{(i)} - 1 + 2\Sigma_+)v_{(i)}\:, \label{vieq}\\
\Omega_{(i)}^\prime &= (2q - 1 - 3w_{(i)})\Omega_{(i)} + (3w_{(i)} - 1 + 2\Sigma_+)Q_{(i)}v_{(i)}\:,
\label{Omieq}\\
\Omega_{\Lambda}^\prime &= 2(1 + q)\Omega_{\Lambda}\:.\label{lambdaeq}
\end{align}
\end{subequations}

\vspace*{2mm} \noindent {\em Constraint equations}:
\begin{subequations} \label{constrBI}
\begin{align}
0 &= 1 -\Sigma^2 - \Omega_{(1)} - \Omega_{(2)} -\Omega_{\Lambda}
\:,\label{gausssys}\\
0 &= Q_{(1)} + Q_{(2)}\label{codazzisys}\:,
\end{align}
\end{subequations}
where
\be\lb{qeq} q = 2\Sigma^2 + \sfrac{1}{2}(\Omega_{\rm m}  +
3P_{\rm m}) -\Omega_{\Lambda} = 2 - \sfrac{3}{2}(\Omega_{\rm m}
- P_{\rm m}) -3\Omega_{\Lambda}\:; \quad \Omega_{\rm m} =
\Omega_{(1)} + \Omega_{(2)}\:, \quad P_{\rm m} = P_{(1)} +
P_{(2)}\:. \ee
Equations~\eqref{vieq} and~\eqref{Omieq} were obtained by using
that~\eqref{dmattereq} takes the same form for non-interacting
individual matter components, where, however, the total matter
content enters into $q$, together with the type I conditions
and the relations~\eqref{pfrel} for the individual perfect
fluids. The assumption of non-negative energy densities and a
non-negative cosmological constant,
$\Omega_{(i)},\Omega_\Lambda \geq 0$, together with~\eqref{qeq}
and~\eqref{gausssys}, yields that $-1\leq q \leq 2$, and hence
that $2-q\geq 0$, where $q=-1$ only when $\Omega_\Lambda=1,
\Omega_m=0,\Sigma^2=0$. It follows that $\tau \in
(-\infty,\infty)$ and $H \rightarrow \infty$ when
$\tau\rightarrow -\infty$ (if $\Omega_\Lambda\neq 1$
initially).

The auxiliary equation,
\be\label{rhoidot} \rho_{(i)}^\prime =
-(1+w_{(i)})(G_+^{(i)})^{-1}[3 + v_{(i)}^2 - 2\Sigma_+
v_{(i)}^2]\rho_{(i)}\:,\ee
implies that $\rho_{(i)}$ is a monotonically decreasing
function such that $\rho_{(i)}\rightarrow \infty$
($\rho_{(i)}\rightarrow 0$) when $\tau\rightarrow -\infty$
($\tau\rightarrow \infty$);
%$Q_{(i)}^\prime &= 2(q - 1 + \Sigma_+) \, Q_{(i)}$
hence the models begin with an initial curvature singularity,
where $\Lambda$ becomes negligible when compared to
$\rho_{(i)}$ when $\tau\rightarrow -\infty$, and then expand
forever to a state where the ordinary matter is infinitely
diluted, leading to that $\Omega_{\rm m}$ becomes negligible
compared to $\Omega_\Lambda$.

%%%%%%%%%%%%%%%%%%%%%%%%%%%%%%%%%%%%%%%%%%%%%%%%%%%%%%%%%%%%%%%%%%%
\section{State space properties}\label{Sec:statprop}
%%%%%%%%%%%%%%%%%%%%%%%%%%%%%%%%%%%%%%%%%%%%%%%%%%%%%%%%%%%%%%%%%%%

%------------------------------------------------------------------
\subsection{The state space}
%------------------------------------------------------------------

The reduced state space consists of ${\bf S} = \{
\Sigma_+,\Sigma_A,\Sigma_B,\Sigma_C,v_{(1)},v_{(2)},
\Omega_{(1)},\Omega_{(2)}, \Omega_{\Lambda}\}$, subject to the
two constraints~\eqref{constrBI}, i.e., the state space is
seven-dimensional. From the definitions and the
constraints~\eqref{constrBI} it follows that the state space is
bounded. Our primary concern in this paper is the `interior'
state space for the case of two {\em tilted\/} fluids for which
$\Omega_{(1)}\Omega_{(2)}>0,\,0<|v_{(1)}v_{(2)}|<1$, however,
the solutions belonging to the interior state space often
asymptotically approach its boundary. To understand the
interior dynamics we therefore consider the closure of the
interior state space, $\bar{{\bf S}}$, thus obtaining a compact
state space, which is possible because of the regularity of the
evolution equations. Hence $\Sigma^2 \leq 1, 0\leq
v_{(i)}^2\leq 1$; $0\leq\Omega_{(1)}\leq 1,
0\leq\Omega_{(2)}\leq 1, 0\leq\Omega_\Lambda\leq 1$, in such a
way so that the constraints~\eqref{constrBI} are satisfied;
note that the Codazzi constraint~\eqref{codazzisys} leads to
that $v_{(1)}v_{(2)}\leq 0$ when $\Omega_{(1)}\Omega_{(2)}>0$,
a condition on $v_{(i)}$ that we extend to the boundary. The
dynamical system~\eqref{evolBI},~\eqref{constrBI} is invariant
under the following discrete symmetries:
\be\label{discrete} \Sigma_A \rightarrow -\Sigma_A\:,\quad
\Sigma_C \rightarrow -\Sigma_C\:;\qquad (v_{(1)},v_{(2)})
\rightarrow -(v_{(1)},v_{(2)})\:. \ee
We therefore assume without loss of generality that $\Sigma_A
\in [0,1], \, \Sigma_C \in [0,1], \, v_{(1)} \in [0,1]$, and
$v_{(2)} \in [-1,0]$; the solutions in the other sectors of the
state space are easily obtained by means of the discrete
symmetries.

%------------------------------------------------------------------
\subsection{The influence of a cosmological constant}
%------------------------------------------------------------------

Equation~\eqref{lambdaeq} implies that $\Omega_\Lambda$ is
monotonically increasing from zero to one. Hence
\begin{equation} \label{Wald}
\Omega_{\Lambda} \rightarrow 1, \quad \Sigma^2 \rightarrow 0, \quad
\Omega_{(i)} \rightarrow 0 \quad \text{when} \quad \tau\rightarrow
\infty\:,
\end{equation}
as follows from combining $\Omega_{\Lambda} \rightarrow 1$ with
the Gauss constraint~\eqref{gausssys}, i.e., the solutions
approach a de Sitter state when $\tau\rightarrow \infty$. This
result is a special case of the proof by Wald~\cite{wal83},
which holds for Bianchi types I-VIII. In the present case the
fluids behave as test fields on a de Sitter background at late
times, obeying the equations: $v_{(i)}^\prime =
(G^{(i)}_-)^{-1}(1-v_{(i)}^2)(3w_{(i)} - 1) \,v_{(i)}$. It
follows that $v_{(i)}=const$ if $w_{(i)}=\frac{1}{3}$;
$v_{(1)}$ is monotonically increasing (decreasing) from 0 to 1
(1 to 0) if $w_{(1)}>\frac{1}{3}$ ($w_{(1)}<\frac{1}{3}$);
$v_{(2)}$ is monotonically decreasing (increasing) from 0 to
$-1$ ($-1$ to 0) if $w_{(2)}>\frac{1}{3}$
($w_{(2)}<\frac{1}{3}$). Thus if one of the fluids has a soft
equation of state, $w_{(2)}<\frac{1}{3}$, and the other has a
sufficiently stiff equation of state, $w_{(1)}\geq
\frac{1}{3}$, then the fluids will obtain a relative velocity
w.r.t each other (in general when $w_{(1)}=\frac{1}{3}$ and
always if $w_{(1)}>\frac{1}{3}$); this is an invariant
statement, and it is not possible to eliminate this effect with
any choice of reference congruence---if one has two fluids, one
with a sufficiently soft and one with a sufficiently stiff
equation of state, then it follows that the fluids will
asymptotically form anisotropies on a de Sitter background
irrespectively of the choice of reference congruence. We note
that this result is compatible with the analysis of Bianchi
type V in~\cite{golnil00}, and that it reflects a bifurcation
that takes place at $w=\frac{1}{3}$ for a fluid in any,
homogeneous or inhomogeneous, forever expanding model with a
cosmological constant, see~\cite{limetal04}.

At early times $\Lambda$ has a negligible effect compared to
normal matter and hence it suffices to study the
$\Omega_\Lambda=0$ subset (it follows from~\eqref{lambdaeq} and
the application of the monotonicity principle, see
e.g.~\cite{waiell97,heiugg06} and references
therein,\footnote{The monotonicity principle gives information
about the global asymptotic behavior of the dynamical system.
If $M: X\rightarrow \mathbb{R}$ is a ${\mathcal C}^1$ function
which is strictly decreasing along orbits (solutions) in $X$,
then $$\omega(x) \subseteq \{\xi \in \bar{X}\backslash X\:|\:
\lim\limits_{\zeta\rightarrow \xi} M(\zeta) \neq
\sup\limits_{X} M\}\:, \quad \alpha(x) \subseteq \{\xi \in
\bar{X}\backslash X\:|\:\lim\limits_{\zeta\rightarrow \xi}
M(\zeta) \neq \inf\limits_{X} M\}$$ for all $x\in X$, where
$\omega(x)$ [$\alpha(x)$] is the $\omega$-limit
[$\alpha$-limit] set of a point $x\in X$, defined as the set of
all accumulation points of the future [past] orbit of $x$; and
analogously for strictly increasing monotonic functions.} that
the $\alpha$-limit for all orbits (solutions) must reside on
this subset (assuming that $\Omega_\Lambda\neq 1$ initially);
cf. also the discussion after equation~\eqref{rhoidot}). Since
$\Lambda$ therefore has no effect on the past asymptotic
dynamics and since it is of interest to also study late time
behavior when one does not have a cosmological constant, we
will from now on assume $\Lambda=0$. The state space we
henceforth therefore consider is given by
\be \bar{\bf S} = \{
\Sigma_+,\Sigma_A,\Sigma_B,\Sigma_C,v_{(1)},v_{(2)},
\Omega_{(1)},\Omega_{(2)}\}\:,
\ee
subject to the constraints~\eqref{constrBI}, i.e., the state
space when one does not have a cosmological constant is
six-dimensional; since the discrete symmetries~\eqref{discrete}
still hold we continue to assume that $\Sigma_A \in [0,1], \,
\Sigma_C \in [0,1], \, v_{(1)} \in [0,1]$, and $v_{(2)} \in
[-1,0]$. When $\Lambda=0$ the deceleration parameter $q$ is
given by
\be q = 2\Sigma^2 + \sfrac{1}{2}(\Omega_{\rm m}  + 3P_{\rm m})
= 2 - \sfrac{3}{2}(\Omega_{\rm m}  - P_{\rm m}) \qquad
\Rightarrow\qquad \sfrac{1}{2} \leq q \leq 2\:. \ee
%

%------------------------------------------------------------------
\subsection{Invariant subsets}
%------------------------------------------------------------------

The dynamical system~\eqref{evolBI},~\eqref{constrBI}, with
$\Omega_\Lambda=0$, admits a number of invariant subsets,
conveniently divided into three classes: (i) `geometric
subsets', i.e., sets associated with conditions on the shear
and hence the metric since the type I models are intrinsically
flat; (ii) invariant sets on the boundary of the physical state
space for two tilted fluids that do not belong to (i); (iii)
subsets that can be obtained by intersections of the subsets
belonging to (i) and (ii). We will introduce a notation where
the kernel suggests the type of subset and where a subscript,
when existent, suggests the values of $v_{(1)}$ and $v_{(2)}$.

\hspace{1mm}

{\em Geometric subsets}
\begin{itemize}
\item ${\cal TW}$: The `twisting' subset, characterized by
    $\Sigma_C=0,\, \Sigma_{A}\neq 0$, which leads to that
    the decoupled $\phi$-variable satisfies $\phi=const$
    and hence $\Sigma_{12} \propto\Sigma_{11}-\Sigma_{22}$.
\item ${\cal RD}$: The constantly rotated diagonal subset,
    given by $\Sigma_A=0,\,\Sigma_C\neq 0$ ($R_\alpha=0$).
    This subset is the diagonal subset, discussed next,
    rotated with a constant angle around $\vece_3$.
\item ${\cal D}$: The diagonal subset, defined by
    $\Sigma_A=\Sigma_C=0;\, \Sigma_B = \Sigma_-$, and hence
    $R_\alpha=0$.
\item ${\cal LRS}$: The locally rotationally symmetric
    subset. This plane symmetric subset of the diagonal
    subset is characterized by the additional condition
    $\Sigma_B=\Sigma_-=0$. This is the simplest subset
    compatible with two tilted fluids.
\item ${\cal FLO}$, ${\cal FLT}_{0v_{(2)}}$and ${\cal
    FLT}_{v_{(1)}0}$: The demand that $\Sigma^2=0$, and
    hence $\Omega_{\rm m}=1$, holds for all times enforces
    either that $v_{(1)}=v_{(2)}=0$, which defines the
    orthogonal Friedmann-Lema\^itre subset ${\cal FLO}$, or
    $v_{(1)}=\Omega_{(2)}=0,\, \Omega_{(1)}=1$
    ($v_{(2)}=\Omega_{(1)}=0,\, \Omega_{(2)}=1$), which
    gives the ${\cal FLT}_{0v_{(2)}}$ (${\cal
    FLT}_{v_{(1)}0}$) Friedmann-Lema\^itre subset with one
    orthogonal fluid and a test vector field $v_{(2)}$
    ($v_{(1)}$); these subsets belong to the boundary of
    the two tilted fluid case and thus there exists no
    Friedmann-Lema\^itre subset with two tilted fluids.
\end{itemize}

\hspace{1mm}

{\em Boundary subsets}
\begin{itemize}
\item ${\cal O}$: The orthogonal subset for which
    $v_{(1)}=v_{(2)}=0$. In general this subset is
    expressed in a non-Fermi frame for which
    $\Sigma_A,\Sigma_C\neq 0$, however, usually when
    dealing with this case one makes a rotation to a Fermi
    frame in which the shear and the metric are diagonal so
    that ${\cal O}$ belongs to ${\cal D}$.
\item ${\cal OT}_{v_{(1)}0}$ and ${\cal OT}_{0v_{(2)}}$:
    The ${\cal OT}_{v_{(1)}0}$ subset describes a single
    orthogonal fluid, $\Omega_{(2)} \geq 0,\ v_{(2)}=0$,
    and a test vector field $v_{(1)}$ ($\Omega_{(1)}=0$),
    and similarly for ${\cal OT}_{0v_{(2)}}$.
\item ${\cal ET}_{1v_{(2)}}$ and ${\cal ET}_{v_{(1)}1}$:
    The subset ${\cal ET}_{1v_{(2)}}$ describes a fluid
    that consists of particles with zero rest mass moving
    with the speed of light $v_{(1)}=1 \,\Rightarrow \,
    Q_{(1)}=\Omega_{(1)} = 3P_{(1)}$, as follows from the
    Codazzi constraint~\eqref{codazzisys}; similar
    statements hold for ${\cal ET}_{v_{(1)}1}$.
\item ${\cal K}$: The vacuum subset is called the Kasner
    subset and is defined by $\Omega_{\rm m}=0;
    \,\Sigma^2=1$; it describes the Kasner solutions, but
    in general in a non-Fermi propagated frame, and with
    $v_{(i)}$ as test fields.
\end{itemize}

The lists above are far from complete; intersections of subsets
are possible in many cases, which then form invariant subsets
of lower dimension; an important example is:
\begin{itemize}
\item ${\cal ET}_{11}={\cal ET}_{1v_{(2)}} \cap {\cal
    ET}_{v_{(1)}1}$: The double extreme tilt subset where
    both fluids propagate with the speed of light,
    $v_{(1)}=1=-v_{(2)}\,\Rightarrow\, \Omega_{(1)} =
    \Omega_{(2)}=3P_{(1)}=3P_{(2)}$, and hence
    $\rho_{(1)}=\rho_{(2)}=3p_{(1)}=3p_{(2)}$.
\end{itemize}

There are also a number of fix points which we denote by a
kernel that is related to a subset to which the fix point
belong together with a subscript that indicates the fix point
values of $v_{(1)}$ and $v_{(2)}$; sometimes we also use a
superscript. The fix points and their stability properties are
given in Appendix~\ref{Sec:locstab}; here we give a brief
summary:
\begin{itemize}
\item There are a number of Kasner points, all satisfying
    $\Omega_m=0,\, \Sigma^2=1,\, q=2$. The four Kasner
    circles: ${\rm K}^{\ocircle}_{00},\, {\rm
    K}^{\ocircle}_{10},\, {\rm K}^{\ocircle}_{01},\, {\rm
    K}^{\ocircle}_{11}$, and the eight Kasner lines: ${\rm
    KL}_{v_{(1)}0}^\pm,\, {\rm KL}_{v_{(1)}1}^\pm,\, {\rm
    KL}_{0v_{(2)}}^\pm,\, {\rm KL}_{1v_{(2)}}^\pm$, where
    the superscript denotes the sign of $\Sigma_B$.
\item There also are a number of Friedmann points with
    $\Sigma^2=0$. The four Friedmann points: ${\rm
    F}^{10}_{00},\, {\rm F}^{10}_{01}$, for which
    $q=\frac{1}{2}(1+3w_{(1)})$, and ${\rm
    F}^{01}_{00},\,{\rm F}^{01}_{10}$, for which
    $q=\frac{1}{2}(1+3w_{(2)})$. When $w_{(2)}=\frac{1}{3}$
    there is a line of fix points, ${\rm
    FL}^{10}_{0v_{(2)}}$, that connects ${\rm F}^{10}_{00}$
    with ${\rm F}^{10}_{01}$, with
    $q=\frac{1}{2}(1+3w_{(1)})$, and similarly when
    $w_{(1)}=\frac{1}{3}$ then ${\rm FL}^{01}_{v_{(1)}0}$
    connects ${\rm F}^{01}_{00}$ with ${\rm F}^{01}_{10}$,
    with $q=\frac{1}{2}(1+3w_{(2)})$. The superscript
    denotes the values of $\Omega_{(1)}$ and
    $\Omega_{(2)}$.
\item When $\frac{1}{3}<w_{(2)}<w_{(1)}$ there exists two
    fix points: ${\rm LRS}_{v_{(1)}^*1}$ with
    $\Sigma^2=\frac{1}{4}(3w_{(1)}-1)^2,\,q=\frac{1}{2}(1+3w_{(1)})$,
    and ${\rm LRS}_{1v_{(2)}^*}$ with
    $\Sigma^2=\frac{1}{4}(3w_{(2)}-1)^2,\,q=\frac{1}{2}(1+3w_{(2)})$.
\item On the twisting subset there exists the extremely
    tilted fix point ${\rm TW}_{11}$, for which
    $\Sigma^2=\frac{2}{5},\,q=\frac{7}{5}$; ${\rm
    TW}_{v_{(1)}^*1}$, which exists when
    $\frac{1}{2}<w_{(1)}<\frac{3}{5}$ with
    $\Sigma^2=\frac{1}{4}(3w_{(1)}-1)(15w_{(1)}-7),\,
    q=\frac{1}{2}(1+3w_{(1)})$; ${\rm TW}_{1v_{(2)}^*}$,
    which exists when $\frac{1}{2}<w_{(2)}<\frac{3}{5}$
    with $\Sigma^2=\frac{1}{4}(3w_{(2)}-1)(15w_{(2)}-7),\,
    q=\frac{1}{2}(1+3w_{(2)})$.
\item Finally there exists the extremely tilted fix point
    ${\rm G}_{11}$ with
    $\Sigma^2=\frac{1}{3},\,q=\frac{4}{3}$; the line of fix
    points ${\rm GL}_{v_{(1)}^*1}$ exists when
    $w_{(1)}=\frac{5}{9}$ with
    $\Sigma^2=\frac{7}{3}v_{(1)}/(3+4v_{(1)}),\,q=\frac{4}{3}$,
    while the line of fix points ${\rm GL}_{1v_{(2)}^*}$
    exists when $w_{(2)}=\frac{5}{9}$ with
    $\Sigma^2=\frac{7}{3}|v_{(2)}|/(3+4|v_{(2)}|),\,q=\frac{4}{3}$.
\end{itemize}
%

%%%%%%%%%%%%%%%%%%%%%%%%%%%%%%%%%%%%%%%%%%%%%%%%%%%%%%%%%%%%%%%%%%%
\section{Monotone functions and their consequences}\label{Sec:mon}
%%%%%%%%%%%%%%%%%%%%%%%%%%%%%%%%%%%%%%%%%%%%%%%%%%%%%%%%%%%%%%%%%%%

In the analysis of Bianchi type VI$_0$ in~\cite{colher04} a
monotone function is defined:
\begin{subequations}
\begin{align}
\chi &= \frac{\beta_{(2)}\Omega_{(2)} -
\beta_{(1)}\Omega_{(1)}}{\beta_{(2)}\Omega_{(2)} +
\beta_{(1)}\Omega_{(1)}} \qquad \text{where} \qquad \beta_{(i)}
= (G_+^{(i)})^{-1} (1-v_{(i)}^2)^{\frac{1}{2}(1-w_{(i)})}\:,\\
\chi^\prime &= \sfrac{3}{2}(w_{(1)} - w_{(2)})(1-\chi^2)\:,
\qquad\qquad\qquad -1\leq \chi \leq 1\:.\lb{chi}
\end{align}
\end{subequations}
The above holds whether or not we include a cosmological
constant. If $w_{(1)} = w_{(2)}$, then $\chi$ is a constant of
the motion, however, here our concern is with the case $w_{(1)}
> w_{(2)}$, and then $\chi$ is a monotonic function that
increases from $-1$ to $1$, which leads to:
\begin{align}
\lim_{\tau\rightarrow -\infty}\, \chi &= -1\quad \Rightarrow\quad
\lim_{\tau\rightarrow -\infty}\,
(\beta_{(2)}\Omega_{(2)}/\beta_{(1)}\Omega_{(1)})=0\quad
\Rightarrow\quad \text{at early times\/}\quad \beta_{(2)}\Omega_{(2)}
\rightarrow 0\:,\label{chicondpast}\\
\lim_{\tau\rightarrow \infty}\, \chi &= 1\quad\,\,\,\, \Rightarrow\quad
\lim_{\tau\rightarrow \infty}\,\,\,\,\:
(\beta_{(1)}\Omega_{(1)}/\beta_{(2)}\Omega_{(2)})=0\quad
\Rightarrow \quad \text{at late times\/}\quad\,\,\, \beta_{(1)}\Omega_{(1)}
\rightarrow 0\:.\label{chicondfuture}
\end{align}
Combined with the Codazzi constraint~\eqref{codazzisys} this
leads to the following possibilities if $\tau\rightarrow
-\infty$:
\begin{itemize}
\item[(i)] $\lim_{\tau\rightarrow
    -\infty}(\Omega_{(1)},\Omega_{(2)})=(0,0)$, i.e., the
    solutions $\alpha$-limits reside on ${\cal K}$,
\item[(ii)] $\lim_{\tau\rightarrow
    -\infty}(\Omega_{(2)},v_1)=(0,0)$, i.e., the solutions
    $\alpha$-limits reside on ${\cal OT}_{0v_{(2)}}$,
\item[(iii)] $\lim_{\tau\rightarrow -\infty}v_{(2)}= -1$,
    $\lim_{\tau\rightarrow -\infty}Q_{(1)}=\Omega_{(2)}$,
    i.e., the solutions $\alpha$-limits reside on ${\cal
    ET}_{v_{(1)}1}$,
\end{itemize}
or combinations/intersections thereof. If $\tau\rightarrow
\infty$ then:
\begin{itemize}
\item[(i)] $\lim_{\tau\rightarrow
    \infty}(\Omega_{(1)},\Omega_{(2)})=(0,0)$, i.e., the
    solutions $\omega$-limits reside on ${\cal K}$,
\item[(ii)] $\lim_{\tau\rightarrow
    \infty}(\Omega_{(1)},v_2)=(0,0)$, i.e., the solutions
    $\omega$-limits reside on ${\cal OT}_{v_{(1)}0}$,
\item[(iii)] $\lim_{\tau\rightarrow \infty}v_{(1)}= 1$,
    $\lim_{\tau\rightarrow \infty}Q_{(1)}=\Omega_{(1)}$,
    i.e., the solutions $\omega$-limits reside on ${\cal
    ET}_{1v_{(2)}}$,
\end{itemize}
or combinations/intersections thereof.

Another monotonic function is given by
\be\lb{vmon} V =
v_{(1)}^2(1-v_{(2)}^2)^{1-w_{(2)}}v_{(2)}^{-2}(1-v_{(1)}^2)^{-(1-w_{(1)})}\:;\qquad
V^\prime = 6(w_{(1)} - w_{(2)})V\:, \ee
where $V$ asymptotically increases from zero to infinity.
Combining $V$ with $\chi$ to obtain a constant of the motion
leads to $Q_{(1)}/Q_{(2)}=const=-1$, where the latter equality
is imposed by the Codazzi constraint, so unfortunately we
obtain nothing new. However, it follows that when
$\tau\rightarrow -\infty$ ($\tau\rightarrow\infty$) then
$v_{(1)}\rightarrow 0$ or/and $v_{(2)}\rightarrow -1$
($v_{(1)}\rightarrow 1$ or/and $v_{(2)}\rightarrow 0$), i.e.,
these limits also hold in the above ${\cal K}$ cases.

Before giving the next monotonic functions it is useful to give
the following auxiliary equations:
\begin{subequations} \label{auxm}
\begin{align}
Q_{(i)}^\prime &= 2(q-1+\Sigma_+)Q_{(i)}\:,\qquad
T_{(i)}^\prime = 2(2q-1-3w_{(i)})T_{(i)}\:, \qquad
\text{where}\lb{QTi}\\
T_{(i)} &=
Q_{(i)}^2(1-v_{(i)}^2)^{1-w_{(i)}}v_{(i)}^{-2}=
(1+w_{(i)})^2(G_+^{(i)})^{-2}\Omega_{(i)}^2(1-v_{(i)}^2)^{1-w_{(i)}}\:.\lb{Ti}
\end{align}
\end{subequations}
If $\Sigma_A,\Sigma_C\neq 0$ there exist two more monotonic
functions:
\be M_{AC}^{(i)} =
Q_{(i)}^{-12}T_{(i)}^9\Sigma_A^{-8}\Sigma_C^{-4} =
Q_{(i)}^6(1-v_{(i)}^2)^{9(1-w_{(i)})}
v_{(i)}^{-18}\Sigma_A^{-8}\Sigma_C^{-4}\:;\qquad
(M_{AC}^{(i)})^\prime = 6(5-9w_{(i)})M_{AC}^{(i)}\:, \ee
where $M_{AC}^{(i)}$ asymptotically increases from zero to
infinity if $w_{i}<\frac{5}{9}$ while it decreases from
infinity to zero if $w_{i}>\frac{5}{9}$; at $w_{i}=\frac{5}{9}$
$M_{AC}^{(i)}$ is a constant of the motion, reflecting that we
have bifurcations when $w_{i}=\frac{5}{9}$, see
Appendix~\ref{Sec:locstab}. The above four monotonic functions,
$\chi,\,V=T_{(2)}/T_{(1)},\,M_{AC}^{(1)},\,M_{AC}^{(2)}$, can
be combined to yield three constants of the motion, but one of
these is just the Codazzi constraint, so there only are two
independent `non-trivial' constants of the motion; here are two
possible representations of these constants of the motion:
\be C_{AB}=
(M_{AC}^{(1)})^{5-9w_{(2)}}(M_{AC}^{(2)})^{9w_{(1)}-5}=
const\:,\qquad D_{AB}= V^{9w_{(1)}-5}(M_{AC}^{(1)})^{w_{(1)} -
w_{(2)}}=const\:.\ee
In addition to these monotonic functions there also exist
several monotonic functions on the various subsets.

The existence of monotone functions is not coincidental, a fact
that will be discussed elsewhere, but let us here comment on
$\chi$, which is a monotonic function for all Class A models
(i.e., Bianchi models for which $A_\alpha=0$, see
section~\ref{Sec:dynsysder} and e.g.~\cite{waiell97}). Its
existence is a consequence of that $\chi$ is expressible as a
dimensionless ratio of the spatial volume density and the
dimensional constants $\ell_{(i)}$ in class A, where
$\ell_{(i)}$ is related to particle conservation of the $i$:th
fluid, see e.g.~\cite{jan01}. Interestingly there exists one
more constant of the motion for each fluid in Class A, however,
these constants of the motion, together with the constants
$\ell_{(i)}$, only lead to an integral related to the Codazzi
constraint~\eqref{codazzisys}, which therefore, unfortunately,
is of no use. Incidentally, other constants of the motion exist
in class B and hence, based on the above insight, there should
exist a monotonic function also in this case, again related to
particle conservation, but in a more complicated way.

%------------------------------------------------------------------
\subsection{The ${\cal K}$ subset}
%------------------------------------------------------------------

\begin{figure}[h]
\psfrag{a}[cc][cc]{$\Sigma_+$} \psfrag{b}[cc][cc]{$\Sigma_3$}
\psfrag{c}[cc][cc]{$\Sigma_B$} \psfrag{d}[cc][cc]{$(321)$}
\psfrag{e}[cc][cc]{$(231)$} \psfrag{f}[cc][cc]{$(213)$}
\psfrag{g}[cc][cc]{$(123)$} \psfrag{h}[cc][cc]{$(132)$}
\psfrag{i}[cc][cc]{$(312)$} \psfrag{j}[cc][cc]{0}
\psfrag{k}[cc][cc]{${\rm T}_3$}\psfrag{l}[cc][cc]{${\rm T}_2$}
\psfrag{m}[cc][cc]{${\rm T}_1$} \psfrag{n}[cc][cc]{${\rm Q}_3$}
\psfrag{o}[cc][cc]{${\rm Q}_2$}\psfrag{p}[cc][cc]{${\rm Q}_1$}
\centering
        \subfigure[Kasner sectors]{
        \label{Ksectors}
        \includegraphics[height=0.30\textwidth]{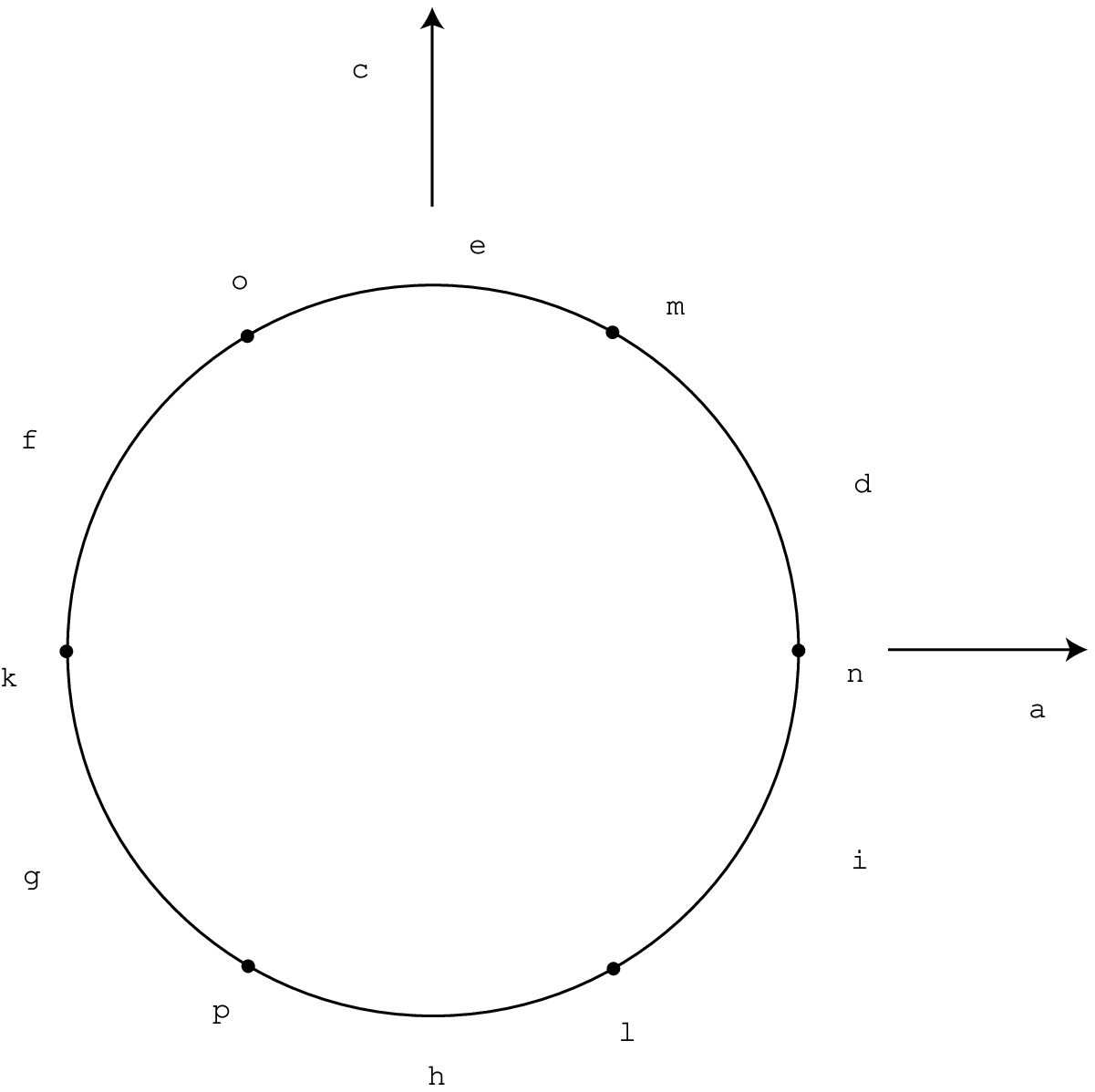}}\qquad
        \subfigure[The $\Sigma_A=0$ subset transitions]{
        \label{KtransC}
        \includegraphics[height=0.30\textwidth]{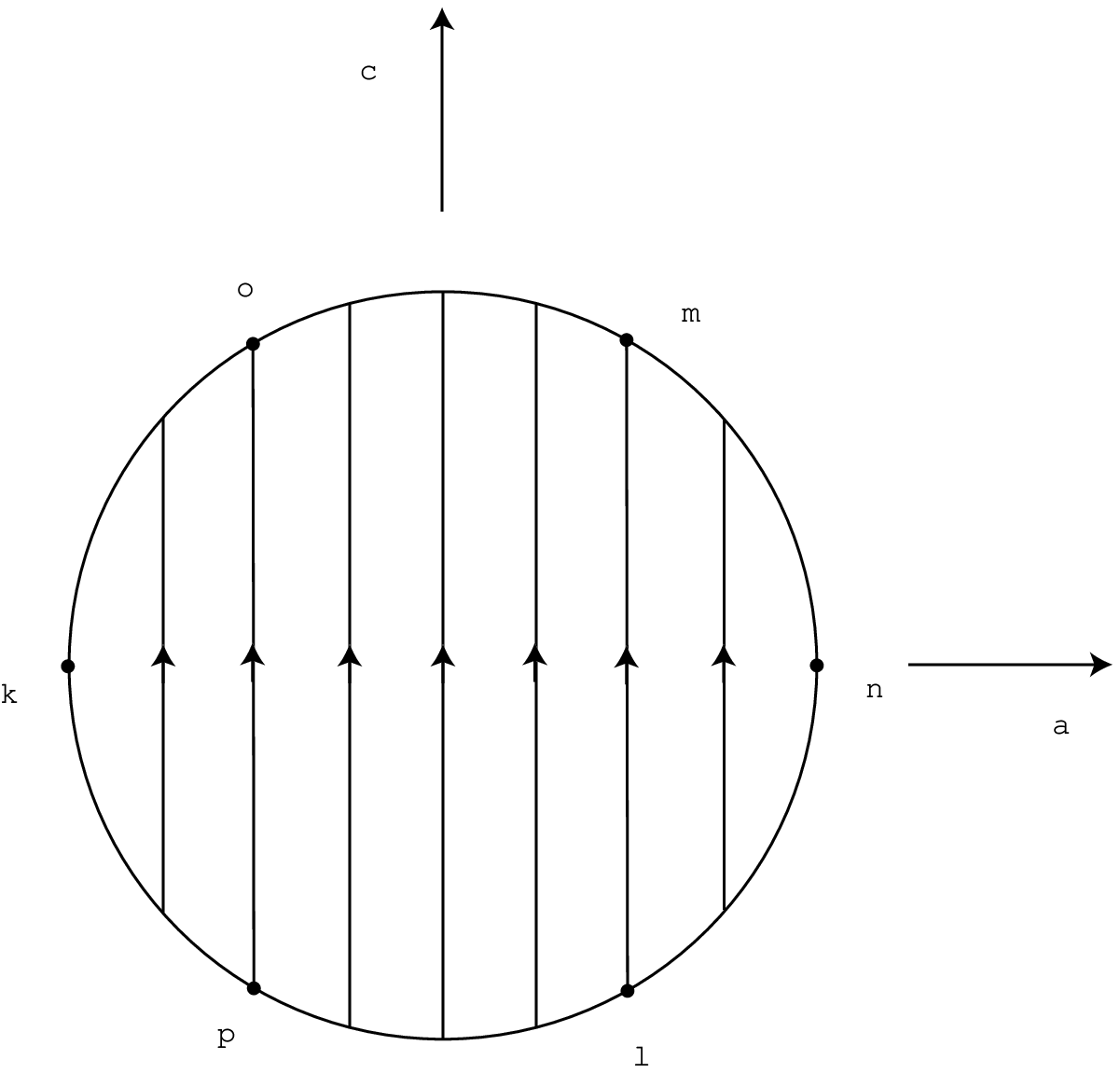}}\qquad
         \subfigure[The $\Sigma_C=0$ subset transitions]{
        \label{KtransA}
        \includegraphics[height=0.30\textwidth]{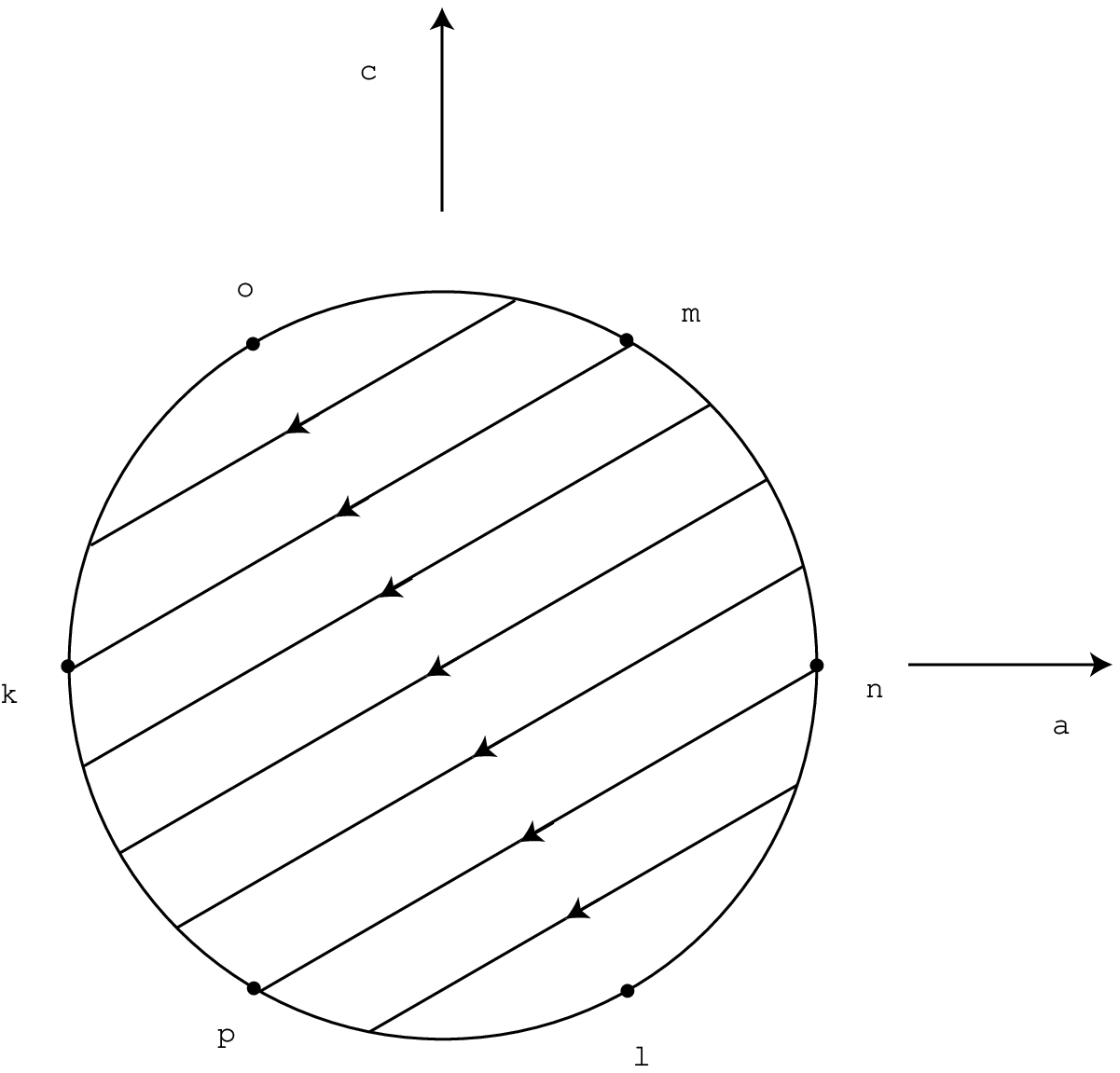}}
        \subfigure[The past attractor segment]{
        \label{Kasnerattractor}
        \includegraphics[height=0.30\textwidth]{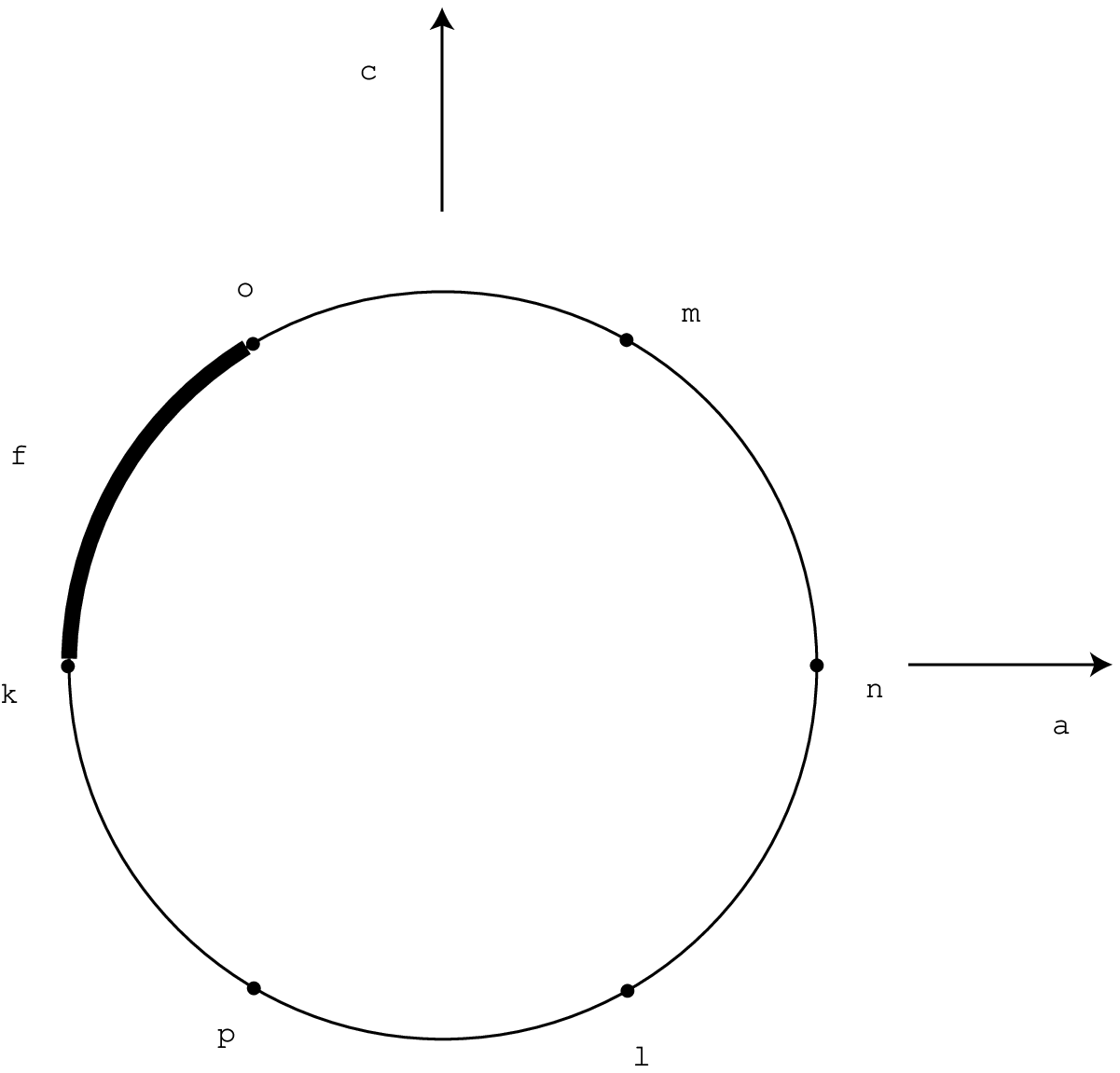}}
        \caption{The projected Kasner circle ${\rm KP}^\ocircle$ is divided into sectors $(i,j,k)$,
        defined by $\Sigma_i<\Sigma_j<\Sigma_k$,
        where $i,j,k$ is a permutation of $1,2,3$, and where
        $\Sigma_1=\hat{\Sigma}_+ +\sqrt{3}\hat{\Sigma}_-$,
        $\Sigma_2=\hat{\Sigma}_+ -\sqrt{3}\hat{\Sigma}_-$,
        $\Sigma_3=-2\hat{\Sigma}_+$, and the points ${\rm Q}_\alpha$,
        corresponding to the non-flat plane symmetric Kasner solution, and ${\rm T}_\alpha$,
        corresponding to the Taub form for the Minkowski spacetime. The Kasner subset is of relevance for
        the past dynamics and hence the arrows point in the {\em past\/} time direction in figures~\ref{KtransC}
        and~\ref{KtransA} which depicts single frame transitions projected onto
        $\Sigma_+-\Sigma_B$-space. The global past attractor
        for the general geometric set with $\Sigma_A\Sigma_B\neq 0$ on ${\cal KP}$
        consists of sector $(213)$ together with ${\rm Q}_2$ and ${\rm T}_3$
        on ${\rm KP}^\ocircle$.}
    \label{Kasner1}
\end{figure}

Before continuing it is useful to discuss the Kasner subset
${\cal K}$. The state space of ${\cal K}$ is given by ${\bf K}
= \{\Sigma_+,\Sigma_A,\Sigma_B,\Sigma_C,v_{(1)},v_{(2)}\}$
subjected to the Gauss constraint
$\Sigma^2=\Sigma_+^2+\Sigma_A^2+\Sigma_B^2+\Sigma_C^2=1$. The
equations for the test fields $v_{(1)}\in [0,1]$, $v_{(2)}\in
[-1,0]$ decouple from those of the shear and from each other.
The state space therefore can be written as the following
Cartesian product:
\be {\bf K}= {\bf KP}\times\{v_{(1)}\}\times\{v_{(2)}\}
\:,\qquad {\bf KP} = \{\Sigma_+,\Sigma_A,\Sigma_B,\Sigma_C\}\:,
\ee
where ${\bf KP}$ is the projected Kasner state space, which of
course is subjected to $\Sigma^2=1$. By determining the
$\alpha$- and $\omega$-limits for solutions on ${\bf KP}$ one
can then determine the asymptotic states of $v_{(1)}$ and
$v_{(2)}$ separately, and thus the $\alpha$- and
$\omega$-limits for solutions on ${\cal K}$. Let us therefore
first turn to the equations on ${\bf KP}$:
\be\label{KP} \Sigma_+^\prime = 3\Sigma_A^2\:;\quad
\Sigma_A^\prime = -(3\Sigma_+ + \sqrt{3}\Sigma_B) \Sigma_A
\:;\quad \Sigma_B^\prime = \sqrt{3}\Sigma_A^2 -
2\sqrt{3}\Sigma_C^2\:;\quad \Sigma_C^\prime =
2\sqrt{3}\Sigma_B\Sigma_C \:. \ee
This system admits a circle of fix points, the projected Kasner
circle: ${\rm KP}^\ocircle$, see Figure~\ref{Ksectors}. It is
described by $\Sigma_A=\Sigma_C=0$,
$\Sigma_+=\hat{\Sigma}_+,\Sigma_B=\hat{\Sigma}_-$, where the
constants $\hat{\Sigma}_\pm$ satisfy $\hat{\Sigma}_+^2 +
\hat{\Sigma}_-^2=1$.

The subset $\Sigma_A=0$ yields that
$\Sigma_+=\hat{\Sigma}_+,\,\Sigma_B^2+\Sigma_C^2=\hat{\Sigma}_-^2$,
where $\Sigma_B$ is monotonically decreasing. The subset
$\Sigma_C=0$ leads to $\Sigma_+ -
\sqrt{3}\Sigma_B=\hat{\Sigma}_+ - \sqrt{3}\hat{\Sigma}_-$ while
$\Sigma_+$ and $\Sigma_B$ are monotonically increasing and
$\Sigma_A^2=1-\Sigma_+^2-\Sigma_B^2$. Projected onto the
$\Sigma_+-\Sigma_B$-plane this yields the straight
lines---single frame transitions, using the nomenclature
of~\cite{heietal07}, given in Figures~\ref{KtransC}
and~\ref{KtransA} (a frame transition preserves a Kasner state
while permuting the spatial axes). As discussed
in~\cite{heietal07}, the general case can be regarded as
multiple frame transitions that yield the same result as
combinations of single transitions which therefore determine
the general asymptotic solution structure on ${\cal KP}$, for
details see~\cite{heietal07}. From this we conclude that the
the $\alpha$-limits for all solutions with
$\Sigma_A\Sigma_C\neq 0$ on ${\cal KP}$ resides on the segment
${\rm KP}^\ocircle$, yielding a segment on ${\rm KP}^\ocircle$
characterized by $-1\leq\Sigma_+=\hat{\Sigma}_+\leq
-\frac{1}{2},\, 0\leq \hat{\Sigma}_-\leq\frac{\sqrt{3}}{2}$,
i.e., the segment consists of sector $(213)$ together with the
fix points ${\rm Q}_2$ and ${\rm T}_3$ on ${\rm KP}^\ocircle$,
see Figure~\ref{Kasnerattractor}.

The $\alpha$-limits for solutions on ${\cal K}$ are determined
by the $\alpha$-limits on ${\cal KP}$ which determine the
asymptotic limits for $v_{(i)}$. The equation for
$|v_{(i)}|\in[0,1]$ on ${\rm KP}^\ocircle$ is given by:
$|v_{(i)}|^\prime = (G^{(i)}_-)^{-1} (1-v_{(i)}^2)(3w_{(i)} - 1
+ 2\hat{\Sigma}_+) \,|v_{(i)}|$. It follows that the
$\alpha$-limits for all orbits on ${\cal K}$ on the general
geometric set with $\Sigma_A\Sigma_C\neq 0$ resides on the
global past attractor ${\cal A}_{\{**\}}$, where the subscript
denotes the range of values of $w_{(1)}$ and $w_{(2)}$, given
by
\begin{subequations}\label{Kattr}
\begin{align}
{\cal A}_{\{w_{(2)}<w_{(1)}<\sfrac{2}{3}\}} &=
\{{\rm K}^\ocircle_{11}: \hat{\Sigma}_+\in [-1,-\sfrac{1}{2}]\}\:,\\
{\cal A}_{\{w_{(2)}<w_{(1)}=\sfrac{2}{3}\}} &=
\{{\rm K}^\ocircle_{11}: \hat{\Sigma}_+\in [-1,-\sfrac{1}{2})\}\cup
\{{\rm KL}^+_{v_{(1)}1}: \hat{\Sigma}_+=-\sfrac{1}{2})\}\:,\\
{\cal A}_{\{w_{(2)}<\sfrac{2}{3}<w_{(1)}\}} &=
\{{\rm K}^\ocircle_{11}: \hat{\Sigma}_+\in [-1,-\sfrac{1}{2}(3w_{(1)}-1)\}\cup
\{{\rm KL}^+_{v_{(1)}1}: \hat{\Sigma}_+=-\sfrac{1}{2}(3w_{(1)}-1)\}\cup\nonumber\\
& \quad\,\, \{{\rm K}^\ocircle_{01}:(-\sfrac{1}{2}(3w_{(1)}-1),-\sfrac{1}{2}]\}\:,\\
{\cal A}_{\{\sfrac{2}{3}=w_{(2)}<w_{(1)}\}} &=
\{{\rm K}^\ocircle_{11}: \hat{\Sigma}_+\in [-1,-\sfrac{1}{2}(3w_{(1)}-1)\}\cup
\{{\rm KL}^+_{v_{(1)}1}: \hat{\Sigma}_+=-\sfrac{1}{2}(3w_{(1)}-1)\}\cup\nonumber\\
& \quad\,\, \{{\rm K}^\ocircle_{01}:(-\sfrac{1}{2}(3w_{(1)}-1), -\sfrac{1}{2})\}
\cup \{{\rm KL}^+_{0v_{(2)}}: \hat{\Sigma}_+= -\sfrac{1}{2}\}\:,\\
{\cal A}_{\{\sfrac{2}{3}<w_{(2)}<w_{(1)}\}} &=
\{{\rm K}^\ocircle_{11}: \hat{\Sigma}_+\in [-1,-\sfrac{1}{2}(3w_{(1)}-1)\}\cup
\{{\rm KL}^+_{v_{(1)}1}: \hat{\Sigma}_+=-\sfrac{1}{2}(3w_{(1)}-1)\}\cup\nonumber\\
& \quad\,\, \{{\rm K}^\ocircle_{01}:(-\sfrac{1}{2}(3w_{(1)}-1),-\sfrac{1}{2}(3w_{(2)}-1))\}
\cup \nonumber\\
& \quad\,\,
\{{\rm KL}^+_{0v_{(2)}}: \hat{\Sigma}_+= -\sfrac{1}{2}(3w_{(2)}-1)\}\cup
\{{\rm K}^\ocircle_{00}: \hat{\Sigma}_+\in(-\sfrac{1}{2}(3w_{(2)}-1),-\sfrac{1}{2}]\}\:.
\end{align}
\end{subequations}

As toward the past, the results in~\cite{heietal07} implies
that all orbits on ${\cal KP}$, on the generic geometric set as
well as all the Kasner compatible geometric subsets,
asymptotically also approach ${\rm KP}^\ocircle$ toward the
future. From this it easily follows from the decoupled
$v_{(i)}$ equations that the $\omega$-limit for any orbit on
${\cal K}$ is one of the Kasner fix points on ${\rm
K}^\ocircle_{**},\, {\rm KL}^\pm_{**}$. But according to the
local stability analysis in Appendix~\ref{Sec:locstab} all fix
points on ${\cal K}$ are destabilized toward the future by the
matter degrees of freedom in the full state space, leading to
that the $\omega$-limit points on ${\cal K}$ become saddles in
the full state space such that no matter solutions with
$Q_{(1)}> 0,\, v_{(1)}^2<1,\,v_{(2)}^2<1$ initially are
attracted to any part of ${\cal K}$ when $\tau\rightarrow
\infty$, and thus the $\omega$-limits for all `interior' matter
solutions either resides on ${\cal OT}_{v_{(1)}0}$ or ${\cal
ET}_{1v_{(2)}}$ such that $q<2$ when $\tau\rightarrow\infty$,
since $q=2$ only on ${\cal K}$.

%%%%%%%%%%%%%%%%%%%%%%%%%%%%%%%%%%%%%%%%%%%%%%%%%%%%%%%%%%%%%%%%%%%
\section{Future and past dynamics}\label{Sec:attractor}
%%%%%%%%%%%%%%%%%%%%%%%%%%%%%%%%%%%%%%%%%%%%%%%%%%%%%%%%%%%%%%%%%%%

%-----------------------------------------------------------------
\subsection{Future dynamics}\label{futattr}
%-----------------------------------------------------------------

The following theorem is easy to prove, but is nevertheless of
interest.
\begin{theorem}\label{noniso}
If $\frac{1}{3}<w_{(2)}<w_{(1)}<1$, and if $Q_{(1)}> 0,\,
v_{(1)}^2<1,\,v_{(2)}^2<1$ initially, then no models isotropize
when $\tau\rightarrow\infty$, i.e., $\Sigma^2\neq 0$ when
$\tau\rightarrow\infty$.
\end{theorem}

\proof Assume that all solutions of the above type isotropize,
i.e. that the $\omega$-limit set for each solution resides on a
Friedmann-Lema\^itre subset. The equations for $v_{(1)}$ and
$v_{(2)}$ then yield $(v_{(1)},v_{(2)})\rightarrow (1,-1)$ when
$\tau\rightarrow\infty$, which is a contradiction since no
Friedmann-Lema\^itre subset has $v_{(1)}v_{(2)}\neq 0$. Hence
none of the solutions described in theorem~\ref{noniso}
isotropize when $\tau\rightarrow\infty$. $\Box$

The above theorem does not tell us where the solutions end up
when $\frac{1}{3}<w_{(2)}<w_{(1)}<1$. This turns out to depend
on what geometric set they belong to, leading to a division of
the models into three classes: (i) The ${\cal RD}$, ${\cal D}$,
${\cal LRS}$ subsets (ii) the ${\cal TW}$ subset, and, (iii)
the general case. Unfortunately we have not been able to prove
what the global attractors are, but our local analysis in
Appendix~\ref{Sec:locstab} together with numerical simulations
lead to the following conjectures:
\begin{conjecture}\label{nonisoid}
If $Q_{(1)}> 0,\, v_{(1)}^2<1,\,v_{(2)}^2<1$ initially, then
the $\omega$-limit for all orbits that belong to the geometric
subsets ${\cal RD}$, ${\cal D}$, and ${\cal LRS}$ is the fix
point ${\rm LRS}_{1v_{(2)}^*}$ if
$\frac{1}{3}<w_{(2)}<w_{(1)}<1$.
\end{conjecture}
\begin{conjecture}\label{nonisoit}
If $Q_{(1)}> 0,\, v_{(1)}^2<1,\,v_{(2)}^2<1$ initially, then
the $\omega$-limit for all orbits that belong to the geometric
subset ${\cal TW}$ ($\Sigma_A\neq 0$) is the fix point ${\rm
LRS}_{1v_{(2)}^*}$ if $\frac{1}{3}<w_{(2)}\leq \frac{1}{2}$ and
$w_{(2)}<w_{(1)}<1$; the fix point ${\rm TW}_{1v_{(2)}^*}$ if
$\frac{1}{2}<w_{(2)}<\frac{3}{5}$ and $w_{(2)}<w_{(1)}<1$; the
fix point ${\rm TW}_{11}$ if $\frac{3}{5}\leq
w_{(2)}<w_{(1)}<1$.
\end{conjecture}
\begin{conjecture}\label{nonisoig}
If $Q_{(1)}> 0,\, v_{(1)}^2<1,\,v_{(2)}^2<1$ initially, then
the $\omega$-limit for all orbits that belong to the general
geometric set ($\Sigma_A,\Sigma_C\neq 0$) is the fix point
${\rm LRS}_{1v_{(2)}^*}$ if $\frac{1}{3}<w_{(2)}\leq
\frac{1}{2}$ and $w_{(2)}<w_{(1)}<1$; the fix point ${\rm
TW}_{1v_{(2)}^*}$ if $\frac{1}{2}<w_{2}<\frac{5}{9}$ and
$w_{(2)}<w_{1}<1$; the line of fix points ${\rm GL}_{1v_{(2)}}$
if $\frac{5}{9}=w_{(2)}<w_{(1)}<1$; the fix point ${\rm
G}_{11}$ if $\frac{5}{9}<w_{(2)}<w_{(1)}<1$.
\end{conjecture}

However, models for which $Q_{(1)}> 0,\,
v_{(1)}^2<1,\,v_{(2)}^2<1$ initially and with $0\leq
w_{(2)}\leq \frac{1}{3}$ {\em do\/} isotropize (this is also
true if $Q_{(1)}=0$, even if the equations of state are stiffer
than radiation), as shown in the following lemma:
\begin{lemma}\lb{liso}
If $Q_{(1)}> 0,\, v_{(1)}^2<1,\,v_{(2)}^2<1$ initially, and if
$0\leq w_{(2)}\leq \frac{1}{3}$, then all models isotropize
when $\tau\rightarrow\infty$, i.e., $\Sigma^2\rightarrow 0$
when $\tau\rightarrow\infty$.
\end{lemma}

\proof In section~\ref{Sec:mon} we showed that the future
$\omega$-limit of a `matter' orbit has to reside on either
${\cal OT}_{v_{(1)}0}$ or ${\cal ET}_{1v_{(2)}}$ with $q<2$ and
$\Omega_{\rm m}>0$. Assume that the $\omega$-limit of an orbit
resides on ${\cal OT}_{v_{(1)}0}$. The equations for the
$\Sigma,\Omega_{(2)}$-variables on this subsets are just those
for a single orthogonal fluid, but in general in a non-Fermi
propagated frame. However, in a Fermi frame the single
orthogonal fluid case is easily solved and one finds that
$\Omega_{(2)}\rightarrow 1$  and $\Sigma^2\rightarrow 0$ when
$\tau\rightarrow\infty$. This statement is frame invariant and
therefore holds for any frame, and hence it follows that the
$\omega$-limit resides on the Friedmann-Lema\^itre subset
${\cal FLT}_{v_{(1)}0}$ and that $\Sigma^2\rightarrow 0$. Let
us now assume that the $\omega$-limit for a matter orbit
resides on ${\cal ET}_{1v_{(2)}}$. Then, since $v_{(1)}=1$,
$q=2 -\Omega_{(1)} -
\frac{3}{2}(\Omega_{(2)}-P_{(2)})=2-\Omega_{\rm m} -
\frac{1}{2}(1-3w_{(2)})(\Omega_{(2)} - Q_{(2)}v_{(2)})$, and
hence $2q-1-3w_{(2)}= 2(1-\Omega_{\rm m})+
(1-3w_{(2)})(1-\Omega_{(2)} + Q_{(2)}v_{(2)})\geq 0$, since
$w_{(2)}\leq\frac{1}{3}$, where the inequality is strict if
$\Omega_{\rm m}<1$, which we now assume. Then $T_{(2)}$
in~\eqref{Ti} is strictly monotonically increasing and grows
without bounds, but this is impossible since $T_{(2)}$ is
finite, and hence $\Omega_{\rm m}\rightarrow 1$, and thus
$\Sigma^2\rightarrow 0$ when $\tau\rightarrow \infty$.$\Box$
\begin{theorem}\label{iso}
If $Q_{(1)}> 0,\, v_{(1)}^2<1,\,v_{(2)}^2<1$ initially, then
the $\omega$-limit for all orbits is the fix point ${\rm
F}_{00}^{01}$ if $0\leq w_{(2)}<w_{(1)}<\frac{1}{3}$; one of
the fix points on the line ${\rm FL}_{v_{(1)}0}^{01}$ if $0\leq
w_{(2)}<w_{(1)}=\frac{1}{3}$; the fix point ${\rm F}_{10}^{01}$
if $0\leq w_{(2)}\leq \frac{1}{3}<w_{(1)}<1$.
\end{theorem}

\proof According to Lemma~\ref{liso} $\Sigma^2=0$
asymptotically toward the future. Imposing this condition on
the ${\cal ET}_{1v_{(2)}}$ subset yields the ${\cal
FLT}_{v_{(1)}0}$ subset with $v_{(1)}=1$, which is a special
case of the other possibility that the $\omega$-limit of an
arbitrary orbit with $Q_{(1)}\neq 0$ initially resides on the
${\cal OT}_{v_{(1)}0}$ subset, and hence that the
$\omega$-limit resides on ${\cal FLT}_{v_{(1)}0}$ with
$v_{(1)}$ so far undetermined ($\Omega_{(2)}=1$). To find the
desired $\omega$-limit we only need to find the asymptotic
limit of $v_{(1)}$, which, according to~\eqref{vieq}, is
determined by the signature of $3w_{(1)}-1$ when $\Sigma^2=0$,
immediately leading to the theorem.

\begin{figure}[h]
\psfrag{a}[cc][cc]{$0$}
\psfrag{b}[cc][cc]{$\frac{1}{3}$}
\psfrag{c}[cc][cc]{$\frac{1}{2}$}
\psfrag{d}[cc][cc]{$\frac{5}{9}$}
\psfrag{e}[cc][cc]{$1$}
\psfrag{f}[cc][cc]{$w_{(1)}$}
\psfrag{g}[cc][cc]{$w_{(2)}$}
\psfrag{h}[cc][cc]{${\rm F}^{01}_{00}$}
\psfrag{i}[cc][cc]{${\rm FL}^{01}_{v_{(1)}0}$}
\psfrag{j}[cc][cc]{${\rm F}^{01}_{10}$}
\psfrag{k}[cc][cc]{${\rm LRS}_{1v^{\ast}_{(2)}}$}
\psfrag{l}[cc][cc]{${\rm TW}_{1v^{\ast}_{(2)}}$}
\psfrag{m}[cc][cc]{${\rm GL}_{1v_{(2)}}$}
\psfrag{n}[cc][cc]{${\rm G}_{11}$}
\psfrag{p}[cc][cc]{${\rm TW}_{11}$}
\psfrag{q}[cc][cc]{$\frac{3}{5}$}
\centering
        \subfigure[The ${\cal RD}$, ${\cal D}$, ${\cal LRS}$ subsets]{
        \label{bifurLRS}
        \includegraphics[height=0.30\textwidth]{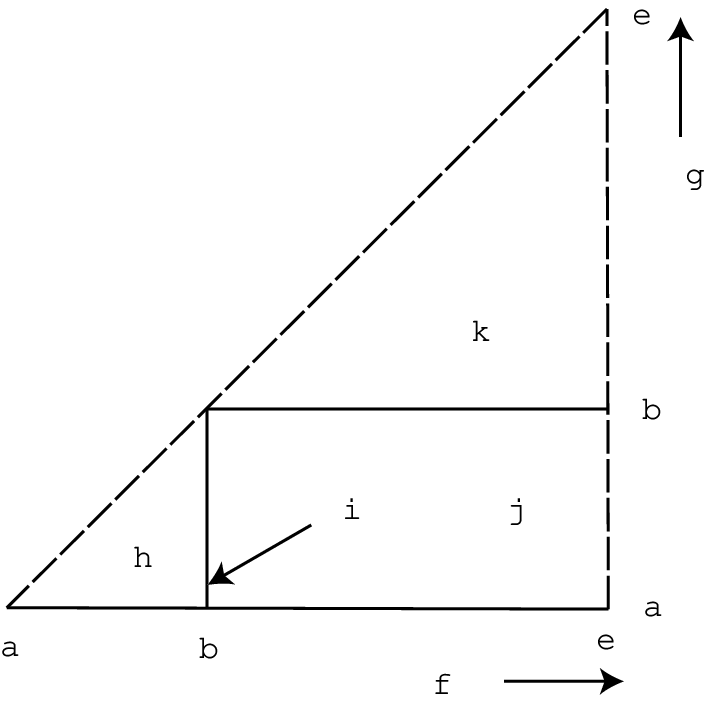}}\qquad
         \subfigure[The ${\cal TW}$ subset]{
        \label{bifurTW}
        \includegraphics[height=0.30\textwidth]{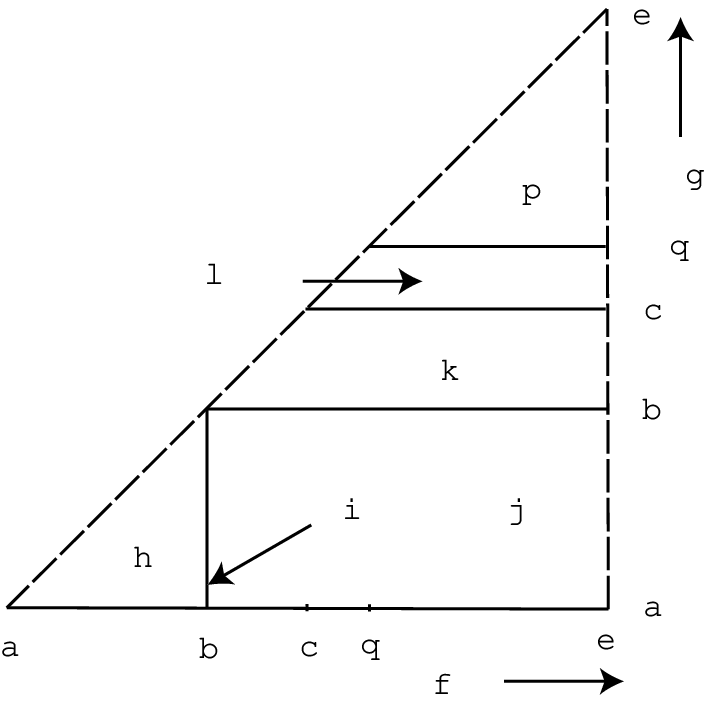}}
        \subfigure[The general case]{
        \label{bifurgen}
        \includegraphics[height=0.30\textwidth]{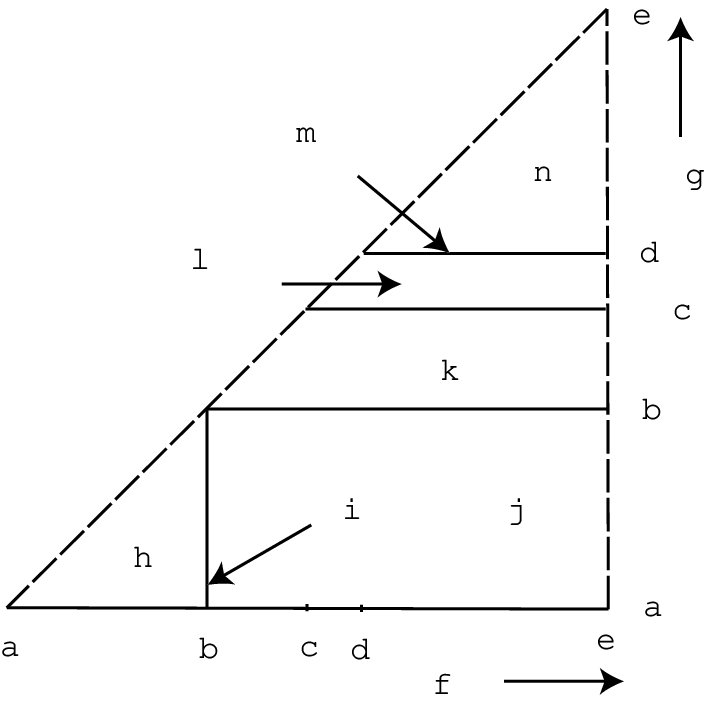}}
        \caption{Future global attractor bifurcation diagrams when $\Lambda=0$ for the
        various geometric subsets.}
    \label{bifur}
\end{figure}

The above theorems and conjectures are summarized in the global
attractor bifurcation diagrams in figure~\ref{bifur}.

%

%-----------------------------------------------------------------
\subsection{Past dynamics}\label{pastattr}
%-----------------------------------------------------------------

Based on the local analysis in Appendix~\ref{Sec:locstab}, the
previous analysis of ${\cal K}$, and a numerical analysis, we
make the following conjecture:
\begin{conjecture}
The $\alpha$-limit for each orbit with $Q_{(1)}> 0,\,
v_{(1)}^2<1,\,v_{(2)}^2<1$ initially on the general geometric
set with $\Sigma_A\Sigma_C\neq 0$ is one of the fix points on
the global past attractor ${\cal A}_{\{**\}}$ for the Kasner
subset ${\cal K}$ given in equation~\eqref{Kattr}.
\end{conjecture}
\begin{remark}
For the various geometric subsets other parts of the projected
Kasner circle are the relevant building blocks for producing
the global attractor for each subset, in a similar way as in
the generic case (e.g., in the ${\cal RD}$ case, with
$\Sigma_A=0,\Sigma_C\neq 0$, $0\leq\hat{\Sigma}_-\leq 1$ is the
restriction on $\hat{\Sigma}_-$, in contrast to the generic
case where $0\leq \hat{\Sigma}_-\leq\frac{\sqrt{3}}{2}$).
\end{remark}
%

%%%%%%%%%%%%%%%%%%%%%%%%%%%%%%%%%%%%%%%%%%%%%%%%%%%%%%%%%%%%%%%%%%%
\section{Concluding remarks}\label{Sec:concl}
%%%%%%%%%%%%%%%%%%%%%%%%%%%%%%%%%%%%%%%%%%%%%%%%%%%%%%%%%%%%%%%%%%%

In this paper we have shown that the type I models with two
tilted fluids exhibit a rich bifurcation structure, hinting at
the complexity one can expect from models with more realistic
sources and more general geometries. Some of our results
reflect features that hold under more general circumstances,
while others are particular for the Bianchi type I models with
two-non-interacting fluids, but in this latter instance the
present models yield a natural reference with which to compare
results from more general settings.

The asymptotically silent regimes of generic spacelike
singularities and of an inflationary future share some
properties: in the inflationary case all other matter fields
than the inflationary field become test fields and do not
influence the spacetime geometry---hence matter that is not
inflationary matter does not matter for the spacetime geometry;
in the case of a generic singularity fluids with speeds of
sound less than that of light also become test fields, in this
case gravity alone creates gravity to a larger extent than
matter, and hence `matter does not matter' in this case
either~\cite{bkl70},~\cite{bkl82},~\cite{uggetal03}. However,
that matter fields asymptotically become test fields does not
mean that they do not matter observationally, on the contrary,
today to a good approximation the CMB can be regarded as a test
field although it is the prime observational source for
cosmology!

In the present case a cosmological constant has yielded a final
de Sitter state---this is a typical feature in a forever
expanding model, as is the bifurcation at the radiation value
$w=\frac{1}{3}$. Hence if one has several test fields, some
less stiff and some as stiff or stiffer than radiation, one
obtains anisotropies on a de Sitter background. However, one
would perhaps not expect fields that are stiffer than radiation
after an inflationary period in the early universe or in the
far future, but does the bifurcation at the radiation value
hint at that e.g. atomic matter and/or cold dark matter and
radiation develop observationally significant relative
velocities, perhaps non-linearly? As regards generic
singularities, the Bianchi type I Kasner singularity is
transformed into a singularity of `Mixmaster' type when one
considers geometrically more general models that admit Bianchi
type II models on the silent boundary in such a combination
with possible frame transitions so that the whole projected
Kasner circle becomes unstable toward the past. But it is by no
means uninteresting to examine the past behavior of type I,
since matter sometimes lead to bifurcations such that matter
sometimes does matter, as illustrated by e.g., a magnetic
field~\cite{leb97}, or by a kinematic description of
matter~\cite{heiugg06} where matter mattered non-generically in
a very subtle way, illustrating that it was not quite obvious
that there would not be any non-generic subtle effects in the
present case; the lack of such effects suggest that the past
dynamics in general is structurally stable under a change from
one to several fluids as long as $0\leq w_{(i)}<1$.

Asymptotic scenarios where non-interacting matter components do
not matter may have interesting consequences when one
introduces more realistic interacting sources. If the
interactions only contribute source terms that are proportional
to the non-interacting parts of the source, then the
interactions presumably also become negligible for the
determination of the geometry; it is only when interactions
contribute more to the total stress-energy than the sources
themselves that the matter does not matter property would be
broken. Hence the approximation of non-interacting fields may
be asymptotically less restrictive than one may initially
think.

There are no (quasi-) isotropic singularities, see
e.g.~\cite{khaetal03},\cite{limetal04} and references therein,
in the present case when $\Sigma^2\neq 0$ initially. The reason
for this is that the shear completely destabilizes such
singularities in Bianchi type I, which therefore is extremely
misleading in a (quasi-) isotropic singularity context.

The result that models with fluids stiffer than radiation
asymptotically produce anisotropies toward the future is
mathematically interesting, and shows that the isotropization
results for a single fluid are structurally unstable within the
Bianchi type I context, although from a physical point of view
one would not expect such equations of states at late times.
The result suggests that tilted fluids may become as
anisotropically significant as spatial curvature at late times
(in the absence of inflation) when one considers more general
models than Bianchi type I, leading to considerable complexity,
further illustrated by the type VI$_0$ investigation
in~\cite{colher04}. Our results about isotropization for soft
equations of state may be regarded as a non-linear Bianchi type
I generalization of perturbations of flat FRW models with two
fluids, a reasonable approximation before dark energy has
becomes significant, and it is of interest then to point out
that one again has radiation bifurcations.

%%%%%%%%%%%%%%%%%%%%%%%%%%%%%%%%%%%%%%%%%%%%%%%%%%%%%%%%%%%%%%%%%%%%%%%%%%
\begin{appendix}
%%%%%%%%%%%%%%%%%%%%%%%%%%%%%%%%%%%%%%%%%%%%%%%%%%%%%%%%%%%%%%%%%%%%%%%%%%

%%%%%%%%%%%%%%%%%%%%%%%%%%%%%%%%%%%%%%%%%%%%%%%%%%%%%%%%%%%%%%%%%%%
\section{Fix points and local stability analysis}\label{Sec:locstab}
%%%%%%%%%%%%%%%%%%%%%%%%%%%%%%%%%%%%%%%%%%%%%%%%%%%%%%%%%%%%%%%%%%%

In this section we use the Gauss constraint~\eqref{gausssys} to
eliminate $\Omega_{(2)}$ globally, however, the Codazzi
constraint~\eqref{codazzisys} cannot, unfortunately, be
analytically solved globally, but we can follow ch.~7
in~\cite{waiell97} and use it to locally eliminate one
variable, usually $\Omega_{(1)}$, at each fix point
when~\eqref{codazzisys} is non-singular. There are several
features that are similar for many of the fix points. All fix
points, except one, have $\Sigma_C=0$; several fix points have
$\Sigma_A=0$. Linearization of~\eqref{SigC} when $\Sigma_C=0$,
and~\eqref{SigA} when $\Sigma_A=0$, yield the eigenvalues
\be \lambda_{\Sigma_C} = -[2 - q_0 -
2\sqrt{3}(\Sigma_B)_0]\:,\qquad \lambda_{\Sigma_A} = -[2 - q_0
+ 3(\Sigma_+)_0 + \sqrt{3}(\Sigma_B)_0]\:, \ee
where $q_0,\, (\Sigma_B)_0,\, (\Sigma_+)_0$ are the fix point
values of $q,\, \Sigma_B,\,\Sigma_+$, respectively. For many
fix points $v_{(i)}=0$ or $|v_{(i)}|=1$. In these cases
linearization of~\eqref{vieq} yields
\be \lambda_{v_{(i)}}^0 = 3w_{(i)} - 1 + 2(\Sigma_+)_0\:,\qquad
\lambda_{v_{(i)}}^1 = -2(3w_{(i)} - 1 +
2(\Sigma_+)_0)/(1-w_{(i)})\:,\ee
where the subscript refers to the $v_{(i)}$ variable the
eigenvalue is connected with while the superscript denotes its
absolute fix point value. Let us now turn to the various
individual fix points; throughout kernel subscripts give an
indication of the absolute fix point values for $v_{(1)}$ and
$v_{(2)}$.

\hspace{1mm}

\noindent {\bf Kasner fix points\/}: There are four circles of
Kasner points and eight lines of fix points when $0\leq
w_{(2)}< w_{(1)}$. The Kasner circles are characterized by
$\Sigma_+ = \hat{\Sigma}_+,\, \Sigma_B = \hat{\Sigma}_-,\,
\Sigma_A = \Sigma_C = 0,\, \Omega_{(1)} = \Omega_{(2)} = 0$,
where $\hat{\Sigma}_\pm$ are constants that satisfy
$\hat{\Sigma}_+^2 + \hat{\Sigma}_-^2=1$, and the following
values of $v_{(i)}$:
\begin{subequations}
\begin{align} {\rm K}^{\ocircle}_{00}:\quad  v_{(1)} &= v_{(2)} =
0\:,\qquad\qquad\, {\rm K}^{\ocircle}_{10}:\,\,  v_{(1)} = 1\:,\, v_{(2)}
= 0\:,\\
{\rm K}^{\ocircle}_{01}:\quad  v_{(1)} &= 0\:,\,
v_{(2)} = -1\:,\qquad {\rm K}^{\ocircle}_{11}:\,\,  v_{(1)} =
-v_{(2)} = 1\:.
\end{align}
\end{subequations}
The eigenvalues for the four cases are:
\begin{subequations}\label{Kasenrcirc}
\begin{align}
& {\rm K}^{\ocircle}_{00}:\quad  0\:;\quad \lambda_{\Sigma_A}\:;\quad
\lambda_{\Sigma_C}\:;\quad \lambda_{v_{(1)}}^0\:;\quad
\lambda_{v_{(2)}}^0\:;\quad 3(1-w_{(1)})\:; \quad  3(1-w_{(2)})\:,\\
& {\rm K}^{\ocircle}_{10}:\quad  0\:;\quad \lambda_{\Sigma_A}\:;\quad
\lambda_{\Sigma_C}\:;\quad \lambda_{v_{(1)}}^1\:;\quad
\lambda_{v_{(2)}}^0\:;\quad  3(1-w_{(2)})\:,\\
& {\rm K}^{\ocircle}_{01}:\quad  0\:;\quad \lambda_{\Sigma_A}\:;\quad
\lambda_{\Sigma_C}\:;\quad \lambda_{v_{(1)}}^0\:;\quad
\lambda_{v_{(2)}}^1\:;\quad  3(1-w_{(1)})\:,\\
& {\rm K}^{\ocircle}_{11}:\quad  0\:;\quad \lambda_{\Sigma_A}\:;\quad
\lambda_{\Sigma_C}\:;\quad \lambda_{v_{(1)}}^1\:;\quad
\lambda_{v_{(2)}}^1\:;\quad  2(1+\hat{\Sigma}_+)\:,
\end{align}
\end{subequations}
where
\begin{subequations}
\begin{align}
\lambda_{\Sigma_A} &=
-(3\hat{\Sigma}_+ + \sqrt{3}\hat{\Sigma}_-)\:, \qquad
\lambda_{\Sigma_C} = 2\sqrt{3}\hat{\Sigma}_-\:,\\
\lambda_{v_{(i)}}^0 &= 3w_{(i)} - 1 + 2\hat{\Sigma}_+\:,\qquad\,\,
\lambda_{v_{(i)}}^1 = -2(3w_{(i)} - 1
+ 2\hat{\Sigma}_+)/(1-w_{(i)})\:.
\end{align}
\end{subequations}
In the ${\rm K}^{\ocircle}_{00}$ case the Codazzi
constraint~\eqref{codazzisys} is singular and hence it cannot
be locally solved; in all other cases~\eqref{codazzisys} has
been used to eliminate $\Omega_{(1)}$. The zero eigenvalue
corresponds to that one has a one-parameter set of fixed
points. The eight lines of Kasner fix points are characterized
by $\Sigma_A = \Sigma_C = 0,\, \Omega_{(1)} = \Omega_{(2)} =
0,\, \Sigma^2=1$, and
\begin{subequations}\label{Kasnerlines}
\begin{align}
{\rm KL}_{v_{(1)}0}^\pm:\quad \Sigma_+ &= \sfrac{1}{2}(1-3w_{(1)})\:,\quad
\Sigma_B = \pm\sqrt{1-\Sigma_+^2}\:,\quad  0 \leq v_{(1)} \leq 1\:, \quad v_{(2)} = 0\:,\\
{\rm KL}_{v_{(1)}1}^\pm:\quad \Sigma_+ &= \sfrac{1}{2}(1-3w_{(1)})\:,\quad
\Sigma_B = \pm\sqrt{1-\Sigma_+^2}\:, \quad 0 \leq v_{(1)} \leq 1\:,\quad v_{(2)} = -1\:,\\
{\rm KL}_{0v_{(2)}}^\pm:\quad \Sigma_+ &= \sfrac{1}{2}(1-3w_{(2)})\:,\quad
\Sigma_B = \pm\sqrt{1-\Sigma_+^2}\:,\quad v_{(1)} = 0\:, \quad -1 \leq v_{(2)} \leq 0\:,\\
{\rm KL}_{1v_{(2)}}^\pm:\quad \Sigma_+ &= \sfrac{1}{2}(1-3w_{(2)})\:,\quad
\Sigma_B = \pm\sqrt{1-\Sigma_+^2}\:,\quad v_{(1)} = 1\:, \quad -1 \leq v_{(2)} \leq 0\:,
\end{align}
\end{subequations}
where the superscript denotes the sign of $\Sigma_B$. After
eliminating $\Omega_{(1)}$ locally by means of the Codazzi
constraint~\eqref{codazzisys}, the eigenvalues for the eight
Kasner lines are:
\begin{subequations}\label{KL}
\begin{align}
&{\rm KL}_{v_{(1)}0}^\pm:\quad 0\:;\quad 0\:;\quad
\lambda_{\Sigma_A}\:;\quad \lambda_{\Sigma_C}\:;\quad
3(1-w_{(2)})\:;\quad  -3(w_{(1)} - w_{(2)})\:,\\
&{\rm KL}_{v_{(1)}1}^\pm:\quad 0\:;\quad 0\:;\quad
\lambda_{\Sigma_A}\:;\quad \lambda_{\Sigma_C}\:;
\quad 3(1-w_{(1)})\:;\quad
6\frac{w_{(1)} - w_{(2)}}{1-w_{(2)}}\:,\\
&{\rm KL}_{0v_{(2)}}^\pm:\quad 0\:;\quad 0\:;\quad
\lambda_{\Sigma_A}\:;\quad \lambda_{\Sigma_C}\:;\quad
3(1-w_{(1)})\:; \quad 3(w_{(1)} - w_{(2)})\:,\\
&{\rm KL}_{1v_{(2)}}^\pm:\quad 0\:;\quad 0\:;\quad
\lambda_{\Sigma_A}\:;\quad \lambda_{\Sigma_C}\:;\quad
3(1-w_{(2)})\:; \quad  -6\frac{w_{(1)} - w_{(2)}}{1-w_{(1)}}\:,
\end{align}
\end{subequations}
where again $\lambda_{\Sigma_A} = -(3\Sigma_+ +
\sqrt{3}\Sigma_B)\:,\, \lambda_{\Sigma_C} = 2\sqrt{3}\Sigma_B$,
where $\Sigma_+,\, \Sigma_B$ take the fix point values for the
relevant line of fix points. Here one zero eigenvalue
corresponds to that one has a line of fix points while the
second is associated with the existence of a one parameter set
of solutions that are anti-parallel w.r.t. each other on each
side of the line of fix points.

\hspace{1mm}

\noindent {\bf Friedmann fix points\/}: All four Friedmann fix
points satisfy $\Sigma_+=\Sigma_A=\Sigma_B=\Sigma_C=0,\,
\Omega_{(1)}\Omega_{(2)}=0,\, \Omega_{\rm m} = 1,\,
v_{(1)}v_{(2)}=0$. They are distinguished by their
$\Omega_{(i)}$ and $v_{(i)}$ values according to:
\begin{subequations}
\begin{align}
& {\rm F}_{00}^{10}:\quad v_{(1)}=0\:,\quad v_{(2)}=0\:,\qquad
\Omega_{(1)}=1\:,\quad \Omega_{(2)}=0\:,\\
& {\rm F}_{01}^{10}:\quad v_{(1)}=0\:,\quad
v_{(2)}=-1\:,\quad\,
\Omega_{(1)}=1\:,\quad \Omega_{(2)}=0\:,\\
& {\rm F}_{00}^{01}:\quad v_{(1)}=0\:,\quad v_{(2)}=0\:,\qquad
\Omega_{(1)}=0\:,\quad \Omega_{(2)}=1\:,\\
& {\rm F}_{10}^{01}:\quad v_{(1)}=1\:,\quad v_{(2)}=0\:,\qquad
\Omega_{(1)}=0\:,\quad \Omega_{(2)}=1\:,
\end{align}
\end{subequations}
where the superscript refers to the values of $\Omega_{(1)}$
and $\Omega_{(2)}$. The associated eigenvalues are:
\begin{subequations}
\begin{align}
& {\rm F}_{00}^{10}:\quad \lambda_{1,2,3,4} = -\sfrac{3}{2}(1-w_{(1)})\:;\quad
3w_{(2)} - 1\:; \quad 3(w_{(1)}-w_{(2)})\:;\qquad v_{(1)}\quad \text{eliminated}\:,\\
& {\rm F}_{01}^{10}:\quad \lambda_{1,2,3,4} =
-\sfrac{3}{2}(1-w_{(1)})\:;\quad
3w_{(1)}-1\:;\quad \frac{2(1-3w_{(2)})}{1-w_{(2)}}\:;\qquad\,\, \Omega_{(1)}\quad \text{eliminated}\:,\\
& {\rm F}_{00}^{01}:\quad \lambda_{1,2,3,4} =
-\sfrac{3}{2}(1-w_{(2)})\:;
\quad 3w_{(1)} - 1\:; \quad -3(w_{(1)}-w_{(2)})\:;\quad\, v_{(2)}\quad \text{eliminated}\:,\\
& {\rm F}_{10}^{01}:\quad \lambda_{1,2,3,4} = -\sfrac{3}{2}(1-w_{(2)})\:; \quad
3w_{(2)} - 1\:;\quad \frac{2(1-3w_{(1)})}{1-w_{(1)}}\:;\qquad\,\, \Omega_{(1)}\quad \text{eliminated}\:.
\end{align}
\end{subequations}
Here the last entry for each line of fix points refers to the
variable that has been eliminated by means of the Codazzi
constraint~\eqref{codazzisys}. Two of the eigenvalues of
$\lambda_{1,2,3,4}$ refer to $\lambda_{\Sigma_A}$ and
$\lambda_{\Sigma_C}$. If $w_{(2)} = \frac{1}{3}$ there exists a
line of Friedmann points, parameterized by $v_{(2)}$, ${\rm
FL}_{0v_{(2)}}^{10}$, that connects ${\rm F}_{00}^{10}$ and
${\rm F}_{01}^{10}$. Similarly if $w_{(1)} = \frac{1}{3}$ there
exists a line of fix points, ${\rm FL}_{v_{(1)}0}^{01}$, that
connects ${\rm F}_{00}^{01}$ and ${\rm F}_{10}^{01}$. They are
given by
\begin{subequations}
\begin{align}
& {\rm FL}_{0v_{(2)}}^{10}:\quad && v_{(1)}=0\:,\quad v_{(2)}=const\:,\qquad
\Omega_{(1)}=1\:,\quad \Omega_{(2)}=0\:, \quad w_{(2)}=1/3\:,\\
& {\rm FL}_{v_{(1)}0}^{01}:\quad && v_{(1)}=const\:,\quad
v_{(2)}=0\:,\qquad
\Omega_{(1)}=0\:,\quad \Omega_{(2)}=1\:, \quad w_{(1)}=1/3\:.
\end{align}
\end{subequations}
The eigenvalues associated with the two lines are:
\begin{subequations}
\begin{align}
& {\rm FL}_{0v_{(2)}}^{10}:\quad \lambda_{1,2,3,4} = -\sfrac{3}{2}(1-w_{(1)})\:;\quad
3w_{(1)}-1\:;\quad 0\:;\qquad\,\, \Omega_{(1)}\quad \text{eliminated}\:,\\
& {\rm FL}_{v_{(1)}0}^{01}:\quad \lambda_{1,2,3,4} =
-\sfrac{3}{2}(1-w_{(2)})\:;
\quad 3w_{(2)} - 1\:; \quad 0\:;\qquad\, \Omega_{(1)}\quad \text{eliminated}\:.
\end{align}
\end{subequations}
We now turn to fix points for which $0<\Sigma^2<1$.

\hspace{1mm}

\noindent {\bf Fix points on ${\cal LRS}\cap {\cal
ET}_{v_{(1)}1}$ and ${\cal LRS}\cap {\cal ET}_{1v_{(2)}}$\/}:
When $\frac{1}{3} < w_{(2)} < w_{(1)}$ there are two additional
fix points, ${\rm LRS}_{v^{\ast}_{(1)}1}$ and ${\rm
LRS}_{1v^{\ast}_{(2)}}$, which enter the physical state space
via ${\rm F}_{01}^{10}$ and ${\rm F}_{10}^{01}$ when
$w_{(1)}=\frac{1}{3}$, $w_{(2)}=\frac{1}{3}$, respectively, and
move into the ${\cal LRS}$-subset with increasing values of
$w_{(i)}$. In the stiff perfect fluid limit ($w_{(1)}=1$,
$w_{(2)}=1$) the lines merge with the coalesced Kasner lines
${\rm KL}_{v_{(1)}1}^+={\rm KL}_{v_{(1)}1}^-$, ${\rm
KL}_{1v_{(2)}}^+={\rm KL}_{1v_{(2)}}^-$, respectively. The two
fix points are characterized by $\Sigma_A = \Sigma_B = \Sigma_C
= 0$, and:
\begin{subequations}
\begin{align}
& {\rm LRS}_{v^{\ast}_{(1)}1}: \quad \Sigma_+ =-\sfrac{1}{2}(3w_{(1)}-1)\:, \qquad
v_{(1)} =\frac{3w_{(1)}-1}{5w_{(1)}+1}\:,\quad v_{(2)} = -1\:,\nonumber \\
& \Omega_{(1)} = \frac{3(1- w_{(1)})(9w_{(1)}+1)(1+w_{(1)})}{32\,w_{(1)}}\:, \qquad
\Omega_{(2)} = \frac{3(1-w_{(1)})(5w_{(1)}+1)(3w_{(1)}-1)}{32\,w_{(1)}}\:,\\
& {\rm LRS}_{1v^{\ast}_{(2)}}: \quad \Sigma_+ = -\sfrac{1}{2}(3w_{(2)}
- 1)\:, \qquad v_{(1)} =1\:, \qquad
v_{(2)} = -\frac{3 w_{(2)} - 1}{5w_{(2)}+1}\:, \nonumber \\
& \Omega_{(1)} =
\frac{3(1-w_{(2)})(5w_{(2)}+1)(3w_{(2)}-1)}{32\,w_{(2)}}\:,
\qquad \Omega_{(2)} = \frac{3(1- w_{(2)})( 9w_{(2)}+1)(1+w_{(2)})}{32\,w_{(2)}}\:.
\end{align}
\end{subequations}
After eliminating $\Omega_{(1)}$ locally the eigenvalues for
the two LRS-points are:
\begin{subequations}
\begin{align}
{\rm LRS}_{v^{\ast}_{(1)}1}: \quad & \lambda_{\Sigma_A} = 3(2w_{(1)}-1)\:; \quad
\lambda_{\Sigma_B} = \lambda_{\Sigma_C} = -\sfrac{3}{2}(1-w_{(1)})\:;
\quad 6 \frac{w_{(1)}-w_{(2)}}{1-w_{(2)}}\:; \nonumber \\
&-\sfrac{3}{4}(1-w_{(1)}) \left(1 \pm
\sqrt{A(w_{(1)})} \right)\:,\nonumber \\
{\rm LRS}_{1v^{\ast}_{(2)}}: \quad & \lambda_{\Sigma_A} =
3(2w_{(2)}-1)\:; \quad \lambda_{\Sigma_B} = \lambda_{\Sigma_C}
= -\sfrac{3}{2}(1-w_{(2)})\:;
\quad  -6 \frac{w_{(1)}-w_{(2)}}{1-w_{(1)}}\:; \nonumber \\
& -\sfrac{3}{4}(1-w_{(2)})\left(1 \pm
\sqrt{A(w_{(2)})} \right)\:,
\end{align}
\end{subequations}
where ${\rm Re}\, A(w_{(i)})<1$; since the expression for
$A(w_{(i)})$ is rather messy we will refrain from giving it.

\hspace{1mm}

\noindent {\bf Fix point on ${\cal TW}\cap {\cal ET}_{11}$\/}:

\be {\rm TW}_{11}: \,\, \Sigma_+ = -\sfrac{2}{5}\:,\,\,
\Sigma_C = 0\:, \,\, \Sigma_A = \Sigma_B =
\sfrac{\sqrt{3}}{5}\:, \quad v_{(1)}=1\:, \,\, v_{(2)} =
-1\:,\,\, \Omega_{(1)}=\Omega_{(2)}=\sfrac{3}{10}\:. \ee
Local elimination of $\Omega_{(1)}$ by means of the Codazzi
constraint~\eqref{codazzisys} yields the eigenvalues:
\be \lambda_{\Sigma_C}=\sfrac{3}{5}\:; \qquad -\sfrac{3}{5}\:;
\qquad -\sfrac{3}{10}(1 \pm i \sqrt{39})\:;\qquad
\frac{6(3-5w_{(2)})}{5(1-w_{(2)})}\:; \qquad
\frac{6(3-5w_{(1)})} {5(1-w_{(1)})}\:. \ee

\hspace{1mm}

\noindent {\bf Fix points on ${\cal TW}\cap {\cal
ET}_{v_{(1)}1}$ and ${\cal TW}\cap {\cal ET}_{1v_{(2)}}$\/}:
When $\frac{1}{2} < w_{(1)} < \frac{3}{5}$ there exists one
more fix point on ${\cal TW}$: ${\rm TW}_{v_{(1)}1}$. This fix
point comes into existence when the point
LRS$_{v^{\ast}_{(1)}1}$ bifurcate into two points at
$w_{(1)}=\frac{1}{2}$; it then wanders away from ${\cal D}$
when $w_{(1)}$ increases and eventually leaves the physical
state space through ${\rm TW}_{11}$ when $w_{(1)}=\frac{3}{5}$.
Yet another similar fix point exists on ${\cal TW}$ if
$\frac{1}{2} < w_{(2)} < \frac{3}{5}$: ${\rm
TW}_{1v^{\ast}_{(2)}}$. The fix points are characterized by
$\Sigma_C = 0$ and
\begin{subequations}
\begin{align}
{\rm TW}_{v^{\ast}_{(1)}1}: \quad \Sigma_+ &=
-\sfrac{1}{2}(3w_{(1)}-1)\:,\qquad \Sigma_A =
\sqrt{\sfrac{3}{2}(1-w_{(1)})(2w_{(1)}-1)}\:,\qquad
\Sigma_B = \sqrt{3}(2w_{(1)} -1)\:,\nonumber \\
v_{(1)} &= v_{(1)}^*=
\frac{(1-w_{(1)})(15w_{(1)}-7)}{-25w_{(1)}^2+18w_{(1)}-1}\:,
\qquad v_{(2)} = -1\:,\nonumber\\
\Omega_{(1)} &= 1 - \sfrac{1}{4}(3w_{(1)}-1)(15w_{(1)}-7) - B(w_{(1)})\:,\qquad
\Omega_{(2)} = B(w_{(1)})\:,\\
{\rm TW}_{1v^{\ast}_{(2)}}: \quad \Sigma_+ &=
-\sfrac{1}{2}(3w_{(2)}-1)\:,\qquad \Sigma_A =
\sqrt{\sfrac{3}{2}(1-w_{(2)})(2w_{(2)}-1)}\:,\qquad
\Sigma_B = \sqrt{3}(2w_{(2)} -1)\:,\nonumber \\
v_{(1)} &= 1\:,\qquad
v_{(2)} = v_{(2)}^* = -\frac{(1-w_{(2)})(15w_{(2)}-7)}{-25w_{(2)}^2+18w_{(2)}-1}\:,\nonumber \\
\Omega_{(1)} &= B(w_{(2)})\:,\qquad
\Omega_{(2)} = 1 - \sfrac{1}{4}(3w_{(2)}-1)(15w_{(2)}-7) - B(w_{(2)})\:,
\end{align}
where
\be B(w_{(i)})= -\frac{3(1 - w_{(i)})(7 - 15w_{(i)})
(25w_{(i)}^2 - 18w_{(i)} + 1)}{32(5w_{(i)}^2 - 5w_{(i)} +
1)}\:.\ee
\end{subequations}
Local elimination of $\Omega_{(1)}$ yields the following
eigenvalues:
\begin{subequations}
\begin{align}
& {\rm TW}_{v^{\ast}_{(1)}1}:\,  \lambda_{\Sigma_C}=-\sfrac{3}{2}(5-9w_{(1)})\:; \quad
6\frac{w_{(1)}-w_{(2)}}{1-w_{(2)}}\:; \,\, \lambda_{3,4,5,6} =
-\sfrac{3}{4}(1-w_{(1)})\left(1 \pm \sqrt{C_{(1)}
\pm D_{(1)}}\right)\:,\\
& {\rm TW}_{1v^{\ast}_{(2)}}:\, \lambda_{\Sigma_C}= -\sfrac{3}{2}(5-9w_{(2)})\:; \,\,
-6\frac{w_{(1)}-w_{(2)}}{1-w_{(1)}}\:; \,\, \lambda_{3,4,5,6} =
-\sfrac{3}{4}(1-w_{(2)})\left(1 \pm \sqrt{C_{(2)}
\pm D_{(2)}}\right)\:,
\end{align}
\end{subequations}
where $C_{(i)}=C_{(i)}(w_{(i)}), D_{(i)}=D_{(i)}(w_{(i)})$
exhibit quite messy expressions, which we therefore refrain
from giving, such that real parts of the associated eigenvalues
always are negative.

\hspace{1mm}

\noindent {\bf Fix point in the generic geometric manifold\/}:
There exists one fix point ${\rm G}_{11}$ for which all the
off-diagonal components of the shear are non-zero. It thus
exists on the generic `geometric' manifold, but on the `matter
boundary' ${\cal ET}_{11}$ where both fluids are extremely
tilted. It is characterized by:
\be {\rm G}_{11}: \,\, \Sigma_+  = -\sfrac{1}{3}\:,\,\,
\Sigma_A = \sfrac{2}{3\sqrt{3}}\:, \,\, \Sigma_B = \Sigma_C
=\sfrac{1}{3\sqrt{3}}\:,\quad v_{(1)}= 1\:,\,\, v_{(2)} =
-1\:,\,\, \Omega_{(1)} = \Omega_{(2)} = \sfrac{1}{3}\:. \ee
Local elimination of $\Omega_{(1)}$ yields the eigenvalues:
\be \lambda_{1,2,3,4} = -\sfrac{1}{3}\left(1 \pm i \sqrt{23 \pm
12\sqrt{2}}\right)\:;\qquad
-\frac{2(9w_{(2)}-5)}{3(1-w_{(2)})}\:;\qquad
-\frac{2(9w_{(1)}-5)}{3(1-w_{(1)})}\:. \ee

At $w_{(2)}<w_{(1)}=\frac{5}{9}$
($w_{(2)}=\frac{5}{9}<w_{(1)}$) there exists a line of fix
points, ${\rm GL}_{v_{(1)}1}$ (${\rm GL}_{1v_{(2)}}$),
connecting ${\rm TW}_{v_{(1)}^*1}$ (${\rm TW}_{1v_{(2)}^*}$)
with $G_{11}$; ${\rm GL}_{v_{(1)}1}$  and ${\rm GL}_{1v_{(2)}}$
are given by:
\begin{subequations}
\begin{align}
{\rm GL}_{v_{(1)}1}: \quad \Sigma_+ & = -\sfrac{1}{3}\:,\qquad
\Sigma_A = \frac{1}{3\sqrt{3}}\sqrt{\frac{34v_{(1)}-6}{3+4v_{(1)}}}\:,\qquad
\Sigma_B = \sfrac{1}{3\sqrt{3}}\:, \quad \Sigma_C =
\frac{1}{3\sqrt{3}}\sqrt{\frac{13v_{(1)}-6}{3+4v_{(1)}}}\nonumber \\
\sfrac{6}{13} & \leq v_{(1)} = const \leq 1\:, \qquad v_{(2)} = -1\:, \nonumber \\
\Omega_{(1)} &= \frac{9+5v_{(1)}^2}{3(1+v_{(1)})(3+4v_{(1)})}
\:, \quad
\Omega_{(2)} = \frac{14v_{(1)}}{3(1+v_{(1)})(3+4v_{(1)})}\:,\\
{\rm GL}_{1v_{(2)}}: \quad \Sigma_+ & = -\sfrac{1}{3}\:,\qquad
\Sigma_A = \frac{1}{3\sqrt{3}}\sqrt{\frac{-34v_{(2)}-6}{3-4v_{(2)}}}\:,\qquad
\Sigma_B = \sfrac{1}{3\sqrt{3}}\:, \quad \Sigma_C =
\frac{1}{3\sqrt{3}}\sqrt{\frac{-13v_{(2)}-6}{3-4v_{(2)}}} \nonumber \\
v_{(1)} &= 1\:,\qquad -1\leq v_{(2)} = const\leq -\sfrac{6}{13}\:,\nonumber \\
\Omega_{(1)} &= \frac{-14v_{(2)}}{3(1-v_{(2)})(3-4v_{(2)})}\:,\qquad
\Omega_{(2)} = \frac{9+5v_{(2)}^2}{3(1-v_{(2)})(3-4v_{(2)})}\:.
\end{align}
\end{subequations}
Local elimination of $\Omega_{(1)}$ yields the eigenvalues:
\begin{subequations}
\begin{align}
& {\rm GL}_{v_{(1)}1}: \,\, 0\:; \qquad\,\,\,\,
\frac{2}{3}\frac{5-9w_{(2)}}{1-w_{(2)}}\:; \quad \lambda_{3,4,5,6} =
-\sfrac{1}{3}\left(1 \pm \sqrt{F_{(1)} \pm G_{(1)}}\right)\:,\\
& {\rm GL}_{1v_{(2)}}: \,\, 0\:; \quad\,\,\,\,\,
-\frac{2}{3}\frac{9w_{(1)}-5}{1-w_{(1)}}\:; \quad \lambda_{3,4,5,6} =
-\sfrac{1}{3}\left(1 \pm \sqrt{F_{(2)} \pm G_{(2)}}\right)\:,
\end{align}
\end{subequations}
where $F_{(i)}=F_{(i)}(v_{(i)}), G_{(i)}=G_{(i)}(v_{(i)})$
exhibit quite messy expressions, which we therefore refrain
from giving, such that real parts of the associated eigenvalues
always are negative.

%%%%%%%%%%%%%%%%%%%%%%%%%%%%%%%%%%%%%%%%%%%%%%%%%%%%%%%%%%%%%%%%%%%%%%%%%%%%%%%%%%%%%%%%%%%%%%%%%%%%%%%%%%
\end{appendix}
%%%%%%%%%%%%%%%%%%%%%%%%%%%%%%%%%%%%%%%%%%%%%%%%%%%%%%%%%%%%%%%%%%%%%%%%%%%%%%%%%%%%%%%%%%%%%%%%%%%%%%%%%%


\begin{thebibliography}{99}

\bi{hinetal08} G.~Hinshaw, et el.
\newblock Five-Year Wilkinson Microwave Anisotropy Probe (WMAP)
Observations: Data Processing, Sky Maps, and Basic Results.
\newblock arXiv:0803.0732.

\bi{tegetal06} M.~Tegmark, et al.
\newblock Cosmological Constraints from the SDSS Luminous Red Galaxies
\newblock Phys.\ Rev.\ D {\bf 74}~:~123507 (2006).

\bi{waiell97} J.~Wainwright and G.~F.~R.~Ellis.
\newblock {\em Dynamical systems in cosmology}.
\newblock {C}ambridge {U}niversity {P}ress, Cambridge, (1997).

\bibitem{col03} A.~A.~Coley.
\newblock {\em Dynamical systems and cosmology}.
\newblock (Kluwer Academic Publishers 2003).

\bi{uggetal03} C.~Uggla, H.~van Elst, J.~Wainwright, and
G.~F~.R.~Ellis.
\newblock The past attractor in inhomogeneous cosmology.
\newblock Phys.\ Rev.\ D {\bf 68}~:~103502 (2003).

\bi{andetal05} L.~Andersson, H.~van Elst, W.~C.~Lim, and
C.~Uggla.
\newblock  Asymptotic Silence of Generic Singularities.
\newblock Phys.\ Rev.\ Lett. {\bf 94} 051101 (2005).

\bi{heietal07} J.~M.~Heinzle, C.~Uggla, and N.~R\"ohr.
\newblock The cosmological billiard attractor.
\newblock arXiv:gr-qc/0702141.

\bi{ellels99} G.~F.~R.~Ellis and H.~van Elst.
\newblock Cosmological models (Carg\`{e}se lectures 1998)
in {\em Theoretical and Observational Cosmology}, edited by
Lachi\`{e}ze-Rey M, (Kluwer, Dordrecht, 1999), p. 1
\newblock ({K}luver: Dortrecht) (gr-qc/9812046)

\bi{rohugg05} N.~R\"ohr and C.~Uggla.
\newblock Conformal regularization of Einstein's field equations
\newblock Class.\ Quantum Grav.\ {\bf 22} 3775 (2005). %3775-3787

\bi{colwai92} A.~A.~Coley and J.~Wainwright.
\newblock Qualitative analysis of two-fluid Bianchi
cosmologies.
\newblock Class.\ Quantum\ Grav.\ {\bf 9} 651 (1992) %651-665

\bi{heietal05} J.~M.~Heinzle, N.~R\"ohr, and C.~Uggla.
\newblock Matter and dynamics in closed cosmologies.
\newblock Phys.\ Rev.\ D {\bf 71}~:~083506 (2005).

\bi{golnil00} M.~Goliath and U.~S.~Nilsson.
\newblock Isotropization of two-component fluids.
\newblock J.\ Math.\ Phys.\ {\bf 41} 6906 (2000). %6906-6917

\bi{colher04} A.~A.~Coley and S.~Hervik.
\newblock A Tale of Two Tilted fluids.
\newblock Class.\ Quantum Grav.\ {\bf 21} 4193 (2004). %4193-4208

\bi{waietal99} J.~Wainwright, M.~J.~Hancock, and C.~Uggla
\newblock Asymptotic self-similarity breaking at late times in
cosmology.
\newblock Class.\ Quantum Grav.\ {\bf 16} 2577 (1999). %2577-2598

\bi{leb97} V.~G.~LeBlanc.
\newblock Asymptotic states of magnetic Bianchi I cosmologies.
\newblock Class.\ Quantum Grav.\ {\bf 14} 2281 (1997). %2281-2301.

\bi{heiugg06} J.~M.~Heinzle and C.~Uggla.
\newblock Dynamics of the spatially homogeneous Bianchi type I Einstein-Vlasov
equations.
\newblock Class.\ Quantum Grav. {\bf 23} 3463 (2006). %3463-3490

\bi{bkl70} V.~A.~Belinski\v{\i}, I.~M.~Khalatnikov, and
E.~M.~Lifshitz.
\newblock Oscillatory approach to a singular point
in the relativistic cosmology.
\newblock Adv.\ Phys.\ {\bf 19}, 525 (1970).
%%CITATION = ADPHA,19,525;%%

\bi{bkl82} V.~A.~Belinski\v{\i}, I.~M.~Khalatnikov, and
E.~M.~Lifshitz.
\newblock A general solution of the Einstein equations with a time singularity.
\newblock Adv.\ Phys.\ {\bf 31}, 639 (1982).
%%CITATION = ADPHA,31,639;%%

\bi{lanlif63} L.~D.~~Landau and E.~M.~Lifshitz
\newblock {\em Fluid Mechanics}.
\newblock {O}xford {P}ergamon, (1963).

\bi{friren00} H.~Friedrich, A.~D.~Rendall.
\newblock The Cauchy Problem for the Einstein Equations
\newblock Lect.\ Notes\ Phys. {\bf 540} 127 (2000).%127-224

\bi{wal83} R.~M.~Wald.
\newblock Asymptotic behavior of homogeneous cosmological
models in the precense of a positive cosmological constant.
\newblock Phys.\ Rev.\ D {\bf 28} 2118 (1983).

\bi{limetal04} W.~C.~Lim, H.~van Elst, C.~Uggla, J.~Wainwright.
\newblock Asymptotic isotropization in inhomogeneous cosmology.
\newblock Phys.\ Rev.\ D {\bf 69}~:~103507 (2004).

\bi{jan01} R.~T.~Jantzen.
\newblock Spatially Homogeneous Dynamics: A Unified Picture.
\newblock arXiv:gr-qc/0102035.

\bi{khaetal03} I.~M.~Khalatnikov, A.~Yu.~Kamenshchik,
M.~Martellini, A.~A.~Starobinsky.
\newblock Quasi-isotropic solution of the Einstein equations near a
cosmological singularity for a two-fluid cosmological model.
\newblock JCAP 0303 (2003) 001.

\end{thebibliography}
\end{document}